\documentclass[11pt]{article}
\usepackage{graphicx}
\usepackage[utf8]{inputenc}
\usepackage{charter}
\usepackage{fullpage}
\usepackage{hyperref}
\usepackage{lipsum}
\usepackage[sort&compress,numbers]{natbib}
\hypersetup{hidelinks}
\usepackage[total={6.5in,9in},top=1in,headsep=0.1in,headheight=1in]{geometry}
\usepackage{enumitem,amssymb}
\usepackage[usenames,dvipsnames]{xcolor}
\usepackage{xfrac}
\usepackage{bm}
\usepackage{array}
\usepackage{comment}
\usepackage{url}
\usepackage{hyperref}[colorlinks = true,
            linkcolor = blue,
            urlcolor  = blue,
            citecolor = blue,
            anchorcolor = blue]

\newcommand\pubnumber{Spectroscopic Roadmap}
\newcommand\pubdate{\today}

\def\Title#1{\begin{center} {\LARGE #1 } \end{center}}

\newcommand\pubblock{\rightline{\begin{tabular}{l} \pubnumber\\
         \pubdate \end{tabular}}}
\newenvironment{Abstract}{\begin{quotation} \begin{center}
                       {\bf ABSTRACT}
                       \end{center}\bigskip  }{\end{quotation}}

\def\beq{\begin{equation}}
\def\eeq#1{\label{#1}\end{equation}}
\def\eeqn{\end{equation}}

\newenvironment{Eqnarray}%
   {\arraycolsep 0.14em\begin{eqnarray}}{\end{eqnarray}}
\def\beqa{\begin{Eqnarray}}
\def\eeqa#1{\label{#1}\end{Eqnarray}}
\def\eeqan{\end{Eqnarray}}

\let\bar=\overbar

\def\lsim{\mathrel{\raise.3ex\hbox{$<$\kern-.75em\lower1ex\hbox{$\sim$}}}}
\def\gsim{\mathrel{\raise.3ex\hbox{$>$\kern-.75em\lower1ex\hbox{$\sim$}}}}

\def\del{\partial}
\def\Dslash{\not{\hbox{\kern-4pt $D$}}}
\def\dslash{\not{\hbox{\kern-2pt $\del$}}}
\def\pslash{\not{\hbox{\kern-2pt $p$}}}
\def\ETmiss{\not{\hbox{\kern-4pt $E$}}_T}

\def\Dlr{\mathrel{\raise1.5ex\hbox{$\leftrightarrow$\kern-1em\lower1.5ex\hbox{$D$}}}}

\def\MSB{{\bar{M \kern -2pt S}}}
\def\msb{{\bar{\scriptsize M \kern -1pt S}}}

\def\drb{{\bar{\scriptsize D \kern -1pt R}}}

\newcommand\apjl{{Astrophysical Journal, Letters}}%
\newcommand\apjs{{ApJS}}%
\newcommand\ao{{Appl.~Opt.}}%
\newcommand\aap{{Astronomy and Astrophysics}}%
\newcommand\snowmass{\begin{center}\rule[-0.2in]{\hsize}{0.01in}\\\rule{\hsize}{0.01in}\\
\vskip 0.1in Submitted to the  Proceedings of the US Community Study\\ 
on the Future of Particle Physics (Snowmass 2021)\\ 
\rule{\hsize}{0.01in}\\\rule[+0.2in]{\hsize}{0.01in} \end{center}}

\usepackage{authblk}

\begin{document}

\pubblock
\snowmass
\Title{A Spectroscopic Road Map for Cosmic Frontier: DESI, DESI-II, Stage-5
}

\bigskip 

\vspace*{-0.3cm}
\begin{center}{{\Large \textsc{Principal Authors}}}\\
\vspace*{0.3cm}
{{David J. Schlegel$^{1}$, Simone Ferraro$^{1}$}}, \\
Greg Aldering$^{1}$,
Charles Baltay$^{2}$,
Segev BenZvi$^{3}$,
Robert Besuner$^{1}$,
Guillermo A. Blanc$^{4}$,
Adam S. Bolton$^{5}$,
Ana~Bonaca$^{4}$,
David Brooks$^{6}$,
Elizabeth Buckley-Geer$^{7}$,
Zheng Cai$^{8}$,
Joseph DeRose$^{1}$,
Arjun Dey$^{5}$,
Peter Doel$^{6}$,
Alex Drlica-Wagner$^{7}$,
Xiaohui Fan$^{9}$,
Gaston Gutierrez$^{7}$,
Daniel Green$^{10}$,
Julien Guy$^{1}$, 
Dragan Huterer$^{11}$,
Leopoldo Infante$^{4}$,
Patrick Jelinsky$^{12}$,
Dionysios Karagiannis$^{13}$,
Stephen M. Kent$^{7}$,
Alex G. Kim$^{1}$,
Jean-Paul Kneib$^{14}$,
Juna A. Kollmeier $^{4,15}$,
Anthony Kremin$^{1}$,
Ofer Lahav$^{6}$,
Martin Landriau$^{1}$,
Dustin Lang$^{16}$,
Alexie Leauthaud$^{17}$,
Michael E.~Levi$^{1}$,
Eric V. Linder$^{12}$,
Christophe Magneville$^{18}$,
Paul Martini$^{19}$,
Patrick McDonald$^{1}$,
Christopher J. Miller$^{11}$,
Adam D. Myers$^{20}$,
Jeffrey A. Newman$^{21}$,
Peter E. Nugent$^{1}$,
Nathalie Palanque-Delabrouille$^{1,18}$,
Nikhil Padmanabhan$^{2}$,
Antonella Palmese$^{22,23,1}$,
Claire Poppett$^{12}$,
Jason X. Prochaska$^{17}$,
Anand Raichoor$^{1}$,
Solange Ramirez$^{4}$,
Noah Sailer$^{1,22}$,
Emmanuel Schaan$^{24,25}$,
Michael Schubnell$^{11}$,
Uros Seljak$^{1,22,26}$,
Hee-Jong Seo$^{27}$,
Joseph Silber$^{1}$,
Joshua D. Simon$^{4}$,
Zachary Slepian$^{28}$,
Marcelle Soares-Santos$^{11}$,
Greg Tarl{\'e}$^{11}$,
Monica Valluri$^{29}$,
Noah J. Weaverdyck$^{1}$,
Risa H. Wechsler$^{24,30,25}$,
Martin White$^{1,26,22}$,
Christophe Y\`eche$^{18}$,
Rongpu Zhou$^{1}$ \\
\vspace*{0.6cm}
{{\Large \textsc{Endorsers}}}\\
Douglas Finkbeiner$^{31}$,
Satya Gontcho A Gontcho$^{1}$,
John Moustakas$^{32}$,
Ashley J. Ross$^{19}$,
Lado Samushia$^{33}$,
Dennis Zaritsky$^{9}$
\end{center}


\newcommand{\Amherst}{University of Massachusetts, Amherst, MA 01003 USA}
\newcommand{\ANLHEP}{HEP Division, Argonne National Laboratory, Lemont, IL 60439, USA}
\newcommand{\APC}{Laboratoire Astroparticule et Cosmologie (APC), CNRS/IN2P3, Universit\'e Paris Diderot, 10, rue Alice Domon et Léonie Duquet, 75205 Paris Cedex 13, France}
\newcommand{\ASU}{Arizona State University, Tempe, AZ  85287}
\newcommand{\BenGurion}{Department of Physics, Ben-Gurion University, Be'er Sheva 84105, Israel}
\newcommand{\BNL}{Brookhaven National Laboratory, Upton, NY 11973}
\newcommand{\Brown}{Brown University, Providence, RI 02912}
\newcommand{\Bub}{Boston University, Boston, MA 02215}
\newcommand{\BU}{Boston University, Boston, MA 02215}
\newcommand{\Buffalo}{Department of Physics, University at Buffalo, SUNY Buffalo, NY 14260 USA}
\newcommand{\Caltech}{California Institute of Technology, Pasadena, CA 91125}
\newcommand{\Cardiff}{School of Physics and Astronomy, Cardiff University, The Parade, Cardiff, CF24 3AA, UK}
\newcommand{\Carleton}{Carleton University, K1S 5B6 Ottawa, Canada}
\newcommand{\Carnegie}{The Observatories of the Carnegie Institution for Science, 813 Santa Barbara St., Pasadena, CA 91101, USA}
\newcommand{\Cavendish}{Astrophysics Group, Cavendish Laboratory, J.J.Thomson Avenue, Cambridge, CB3 0HE, UK}
\newcommand{\CCA}{Center for Computational Astrophysics, 162 5th Ave, 10010, New York, NY, USA}
\newcommand{\CITA}{Canadian Institute for Theoretical Astrophysics, 60 St. George St., Toronto Canada}
\newcommand{\CPPM}{Aix Marseille Univ, CNRS/IN2P3, CPPM, Marseille, France}

\newcommand{\CEADAP}{D\'epartement d’Astrophysique, CEA Saclay DSM/Irfu, 91191 Gif-sur-Yvette, France}
\newcommand{\CERN}{CERN, Geneva, Switzerland}
\newcommand{\CfA}{Harvard-Smithsonian Center for Astrophysics, Cambridge, MA 02138}
\newcommand{\CFT}{Center for Theoretical Physics, Polish Academy of Sciences, al. Lotnik\'{o}w 32/46, 02-668, Warsaw, Poland}
\newcommand{\Cincinnati}{University of Cincinnati, Cincinnati, OH 45221}

\newcommand{\CNRSA}{CNRS, Laboratoire d'Annecy-le-Vieux de Physique Th\'{e}orique, France}
\newcommand{\CNYang}{C.N. Yang Institute for Theoretical Physics State University of New York Stony Brook, NY 11794}
\newcommand{\CMUCosmo}{Department 
of Physics, McWilliams Center for Cosmology, Carnegie Mellon University}
\newcommand{\Columbia}{Columbia University, New York, NY 10027}
\newcommand{\Cornell}{Cornell University, Ithaca, NY 14853}
\newcommand{\CPthree}{CP3-Origins, 5230 Odense, Denmark}
\newcommand{\daa}{Department of Astronomy and Astrophysics, University of Toronto, ON, M5S3H4}
\newcommand{\damtp}{DAMTP, Centre for Mathematical Sciences, Wilberforce Road, Cambridge, UK, CB3 0WA}
\newcommand{\DESY}{DESY,  22607 Hamburg, Germany}
\newcommand{\DFI}{Departamento de F\'isica, FCFM, Universidad de Chile, Blanco Encalada 2008, Santiago, Chile}
\newcommand{\DOE}{US. Department of Energy, Germantown, MD 20874}
\newcommand{\drexel}{Drexel University, Philadelphia, PA 19104}
\newcommand{\Duke}{Duke University and Triangle Universitites Nuclear Laboratory, Durham, NC 27708}
\newcommand{\DukePhys}{Department of Physics, Duke University, Durham, NC 27708, USA}
\newcommand{\dunlap}{Dunlap Institute for Astronomy and Astrophysics, University of Toronto, ON, M5S3H4}
\newcommand{\Durham}{Department of Physics, Lower Mountjoy, South Rd, Durham DH1 3LE, United Kingdom}
\newcommand{\ED}{University of Edinburgh, EH8 9YL Edinburgh, United Kingdom}
\newcommand{\EPFL}{Institute of Physics, Laboratory of Astrophysics, Ecole Polytechnique Fédérale de Lausanne (EPFL), Observatoire de Sauverny, 1290 Versoix, Switzerland}
\newcommand{\ETH}{ETH Zurich, Institute for Particle Physics, 8093 Zurich, Switzerland}
\newcommand{\EPFLEng}{STI IMT, École Polytechnique Fédérale de Lausanne (EPFL), 1015 Lausanne, Switzerland}
\newcommand{\FNAL}{Fermi National Accelerator Laboratory, Batavia, IL 60510}
\newcommand{\FQAUB}{Dept. de F\' isica Qu\` antica i Astrof\' isica, Universitat de Barcelona, Mart\' i i Franqu\` es 1, E08028 Barcelona, Spain}
\newcommand{\FSU}{Florida State University, Tallahassee, FL 32306}
\newcommand{\Glasgow}{University of Glasgow, G12 8QQ Glasgow, United Kingdom}
\newcommand{\GRAPPA}{GRAPPA Institute, University of Amsterdam, Science Park 904, 1098 XH Amsterdam, The Netherlands}
\newcommand{\GSFC}{Goddard Space Flight Center, Greenbelt, MD 20771 USA}
\newcommand{\GWU}{George Washington University, Washington, DC 20052}
\newcommand{\Hampton}{Hampton University, Hampton, VA 23668}
\newcommand{\HarvardPhys}{Department of Physics, Harvard University, Cambridge, MA 02138, USA}
\newcommand{\Haverford}{Haverford College, 370 Lancaster Ave, Haverford PA, 19041, USA}
\newcommand{\Hawaii}{University of Hawaii, Honolulu, HI 96822}
\newcommand{\HKUST}{The Hong Kong University of Science and Technology, Hong Kong SAR, China}
\newcommand{\houston}{University of Houston, Houston, TX 77204}
\newcommand{\IAP}{Institut d'Astrophysique de Paris (IAP), CNRS \& Sorbonne University, Paris, France}
\newcommand{\IAS}{Institute for Advanced Study, Princeton, NJ 08540}
\newcommand{\IBS}{Institute for Basic Science (IBS), Daejeon 34051, Korea}
\newcommand{\ICC}{ICC, University of Barcelona, IEEC-UB, Mart\' i i Franqu\` es, 1, E08028 Barcelona, Spain}
\newcommand{\ICCD}{Institute for Computational Cosmology, Department of Physics, Durham University, South Road, Durham, DH1 3LE, UK}
\newcommand{\ICE}{Institute of Space Sciences (ICE, CSIC), Campus UAB, Carrer de Can Magrans, s/n, 08193 Barcelona, Spain}
\newcommand{\ICRR}{Institute for Cosmic Ray Resaerch, The University of Tokyo, 456 Higashi-Mozumi, Kamioka, Hida, Gifu 506-1205, Japan}
\newcommand{\ICTP}{International Centre for Theoretical Physics, Strada Costiera, 11, I-34151 Trieste, Italy}
\newcommand{\IFAE}{Institut de Fisica d’Altes Energies, The Barcelona Institute of Science and Technology, Campus UAB, 08193 Bellaterra (Barcelona), Spain}
\newcommand{\IFPU}{IFPU - Institute for Fundamental Physics of the Universe, Via Beirut 2, 34014 Trieste, Italy}
\newcommand{\IFT}{Instituto de Fisica Teorica UAM/CSIC, Universidad Autonoma de Madrid, 28049 Madrid, Spain}
\newcommand{\IFUNAM}{IFUNAM - Instituto de F\'{i}sica, Universidad Nacional Aut\'onoma de M\'etico, 04510 CDMX, M\'exico}
\newcommand{\IHEP}{Institute of High Energy Physics, Austrian Academy of Sciences, 1050 Vienna, Austria}
\newcommand{\Imperial}{Theoretical Physics, Blackett Laboratory, Imperial College, London, SW7 2AZ, U.K.}
\newcommand{\Indiana}{Indiana University, Bloomington, IN 47405}
\newcommand{\INAFOATs}{INAF - Osservatorio Astronomico di Trieste, Via G.B. Tiepolo 11, 34143 Trieste, Italy}
\newcommand{\INAFOAS}{INAF - Osservatorio di Astrofisica e Scienza dello Spazio di Bologna, via Piero Gobetti 93/3, I-40129 Bologna, Italy}
\newcommand{\INFNCag}{Istituto Nazionale di Fisica Nucleare, Sezione di Cagliari,  09126 Cagliari, Italy}
\newcommand{\INFNCat}{Istituto Nazionale di Fisica Nucleare, Sezione di Catania, 95125 Catania, Italy}
\newcommand{\INFNG}{Istituto Nazionale di Fisica Nucleare, Sezione di Genova, 16146 Genova, Italy}
\newcommand{\INFN}{INFN – National Institute for Nuclear Physics, Via Valerio 2, I-34127 Trieste, Italy}
\newcommand{\INFNLNF}{Istituto Nazionale di Fisica Nucleare, Laboratori Nazionali di Frascati, 00044 Frascati, Italy}
\newcommand{\INFNLNS}{Istituto Nazionale di Fisica Nucleare, Laboratori Nazionali del Sud, 95125 Catania, Italy}
\newcommand{\INFNN}{Istituto Nazionale di Fisica Nucleare, Sezione di Napoli, 80125 Napoli, Italy }
\newcommand{\INFNRM}{Istituto Nazionale di Fisica Nucleare, Sezione di Roma, 00185 Roma, Italy}
\newcommand{\INFNT}{Istituto Nazionale di Fisica Nucleare, Sezione di Torino, 10125, Italy }
\newcommand{\ioa}{Institute of Astronomy, University of Cambridge,Cambridge CB3 0HA, UK}
\newcommand{\IPP}{Institute for Particle Physics, BC V8W 3P6 Victoria, Canada}
\newcommand{\IPMU}{Kavli Insitute for the Physics and Mathematics of the Universe (WPI), University of Tokyo, 277-8583 Kashiwa , Japan}
\newcommand{\IPNL}{Universit\'e de Lyon, F-69622, Lyon, France; Universit\'e de Lyon 1, Villeurbanne; CNRS/IN2P3, Institut de Physique Nucl\'eaire de Lyon}
\newcommand{\IRFU}{IRFU, CEA, Universit\'e Paris-Saclay, F-91191 Gif-sur-Yvette, France}
\newcommand{\ITFA}{Institute for Theoretical Physics, University of Amsterdam, Science Park 904, 1098 XH Amsterdam, The Netherlands}
\newcommand{\IUCAA}{The Inter-University Centre for Astronomy and Astrophysics, Pune, 411007, India}
\newcommand{\Jerusalem}{Hebrew University of Jerusalem, 91904 Jerusalem, Israel}
\newcommand{\JHU}{Johns Hopkins University, Baltimore, MD 21218}
\newcommand{\JLAB}{Thomas Jefferson National Laboratory, Newport News, VA 23606}
\newcommand{\JPL}{Jet Propulsion Laboratory, California Institute of Technology, Pasadena, CA, USA}
\newcommand{\KASSI}{Korea Astronomy and Space Science Institute, Daejeon 34055, Korea}
\newcommand{\kavli}{Kavli Institute for Cosmology, Cambridge, UK, CB3 0HA}
\newcommand{\KIAS}{School of Physics, Korea Institute for Advanced Study, 85 Hoegiro, Dongdaemun-gu, Seoul 130-722, Korea}
\newcommand{\KICP}{Kavli Institute for Cosmological Physics, Chicago, IL 60637}
\newcommand{\KIPAC}{Kavli Institute for Particle Astrophysics and Cosmology, Stanford University and SLAC, Stanford, CA 94305}
\newcommand{\KINGS}{King's College London, WC2R 2LS London, United Kingdom}
\newcommand{\Kobe}{Kobe University, 657-8501 Kobe, Japan}
\newcommand{\KPH}{Johannes Gutenberg University, 55128 Mainz, Germany}
\newcommand{\KPMU}{University of Tokyo, 277-8583  Kashiwa , Japan}
\newcommand{\KSU}{Kansas State University, Manhattan, KS 66506}
\newcommand{\Lafayette}{Lafayette College, Easton, PA 18042}
\newcommand{\LANL}{Los Alamos National Laboratory, Los Alamos, NM 87545}
\newcommand{\LBL}{Lawrence Berkeley National Laboratory, Berkeley, CA 94720}
\newcommand{\Leiden}{Lorentz Institute, Leiden University, Niels Bohrweg 2,Leiden, NL 2333 CA, The Netherlands}
\newcommand{\Liverpool}{University of Liverpool,  L69 7ZE Liverpool , United Kingdom}
\newcommand{\LLNL}{Lawrence Livermore National Laboratory, Livermore, CA, 94550}
\newcommand{\LPC}{Universit\'e Clermont Auvergne, CNRS/IN2P3, Laboratoire de Physique de Clermont, F-63000 Clermont-Ferrand, France}
\newcommand{\LPNHE}{Sorbonne Universit\'e, Universit\'e Paris Diderot, CNRS/IN2P3, Laboratoire de Physique Nucl\'eaire et de Hautes Energies, LPNHE, 4 Place Jussieu, F-75252 Paris, France}
\newcommand{\McGill}{McGill University, Montreal, QC H3A 2T8, Canada}
\newcommand{\Melbourne}{School of Physics, The University of Melbourne, Parkville, VIC 3010, Australia}
\newcommand{\Mines}{Colorado School of Mines, Golden, CO 80401}
\newcommand{\MIT}{Massachusetts Institute of Technology, Cambridge, MA 02139}
\newcommand{\MPE}{Max-Planck-Institut f\"{u}r extraterrestrische Physik (MPE), Giessenbachstrasse 1, D-85748 Garching bei M\"unchen, Germany}
\newcommand{\MPHeidelberg}{Max Planck Institut für Astronomie, Königstuhl 17, D–69117 Heidelberg, Germany}
\newcommand{\MPIA}{Max-Planck-Institut f\"{u}r Astrophysik, Karl-Schwarzschild-Str. 1, 85741 Garching, Germany}
\newcommand{\LUPM}{Laboratoire Univers et Particules de Montpellier, Univ. Montpellier and CNRS, 34090 Montpellier, France}
\newcommand{\NAOC}{National Astronomical Observatories, Chinese Academy of Sciences, PR China}
\newcommand{\NCBJ}{National Center for Nuclear Research, Ul.Pasteura 7,Warsaw, Poland}
\newcommand{\NCU}{National Central University, Taoyuan City 32001, Taiwan (R.O.C.)}
\newcommand{\NCSU}{Physics Department, North Carolina State Universitym, 2401 Stinson Dr, Raleigh, NC 27607}
\newcommand{\ND}{University of Notre Dame,vNotre Dame, IN 46556}
\newcommand{\NIU}{Northern Illinois University, DeKalb, Illinois 60115}
\newcommand{\NMSU}{New Mexico State University, Las Cruces, NM 88003}
\newcommand{\NOAO}{NSF's NOIRLab, 950 N. Cherry Ave., Tucson, AZ 85719 USA}
\newcommand{\Northwestern}{Northwestern University, Evanston, IL 60201}
\newcommand{\Nottingham}{University of Nottingham, NG7 2RD Nottingham, United Kingdom}
\newcommand{\NWU}{Northwestern University, Evanston, IL 60208}
\newcommand{\NYU}{New York University, New York, NY 10003}
\newcommand{\OK}{ University of Oklahoma, Norman, OK 73019}
\newcommand{\ORNL}{Oak Ridge National Laboratory, Oak Ridge, TN 37831}
\newcommand{\OSU}{The Ohio State University, Columbus, OH 43212}
\newcommand{\OU}{Department of Physics and Astronomy, Ohio University, Clippinger Labs, Athens, OH 45701, USA}
\newcommand{\OskarKlein}{Oskar Klein Centre for Cosmoparticle Physics, Stockholm University, AlbaNova, Stockholm SE-106 91, Sweden}
\newcommand{\Oxford}{The University of Oxford, Oxford OX1 3RH, UK}
\newcommand{\Oxy}{Occidental College, Los Angeles, CA 90041}
\newcommand{\ParisSud}{Universit\'{e} Paris-Sud, LAL, UMR 8607, F-91898 Orsay Cedex, France \& CNRS/IN2P3, F-91405 Orsay, France}
\newcommand{\PI}{Perimeter Institute, Waterloo, Ontario N2L 2Y5, Canada}
\newcommand{\Pitt}{University of Pittsburgh and PITT PACC, Pittsburgh, PA 15260}
\newcommand{\PNNL}{Pacific Northwest National Laboratory ,Richland, WA 99352}
\newcommand{\PNPI}{Petersburg Nuclear Physics Institute, 188300 Gatchina, Russia}
\newcommand{\Port}{Institute of Cosmology \& Gravitation, University of Portsmouth, Dennis Sciama Building, Burnaby Road, Portsmouth PO1 3FX, UK}
\newcommand{\Princeton}{Princeton University, Princeton, NJ 08544}
\newcommand{\PSU}{The Pennsylvania State University, University Park, PA 16802}
\newcommand{\Purdue}{Purdue University, West Lafayette, IN 47907}
\newcommand{\PW}{Participation Worldscope, Sedona, Arizona and Hong Kong, SAR PRC}
\newcommand{\Queens}{Queen's University , K7L 3N6 Kingston, Canada}
\newcommand{\Queensland}{The University of Queensland, School of Mathematics and Physics, QLD 4072, Australia}
\newcommand{\QMUL}{Queen Mary University of London, Mile End Road, London E1 4NS, United Kingdom}
\newcommand{\RAL}{Radio Astronomy Laboratory, University of California Berkeley, Berkeley, CA 94720, USA}
\newcommand{\Rice}{Department of Physics & Astronomy, Rice University, Houston, Texas 77005, USA}
\newcommand{\RIT}{Rochester Institute of Technology}
\newcommand{\RomaS}{Dipartimento di Fisica, Universit\`{a} La Sapienza, P. le A. Moro 2, Roma, Italy}
\newcommand{\RUG}{Kapteyn Astronomical Institute, University of Groningen, P.O. Box 800, 9700 AV Groningen, The Netherlands}
\newcommand{\Rutgers}{Department of Physics and Astronomy, Rutgers, the State University of New Jersey, 136 Frelinghuysen Road, Piscataway, NJ 08854, USA}
\newcommand{\Sanford}{Sanford Underground Research Facility, Lead, SD 57754}
\newcommand{\Sassari}{Universit\`a di Sassari, 07100 Sassari,  Italy}
\newcommand{\SCIPP}{University of California at Santa Cruz, Santa Cruz, CA 95064}
\newcommand{\Sejong}{Department of Physics and Astronomy, Sejong University, Seoul, 143-747, Korea}
\newcommand{\Sheffield}{University of Sheffield, S3 7RH Sheffield, United Kingdom}
\newcommand{\SHAO}{Shanghai Astronomical Observatory (SHAO), Nandan Road 80, Shanghai 200030, China}
\newcommand{\Siena}{Siena College, Department of Physics \& Astronomy, 515 Loudon Road, Loudonville, NY 12211, USA}
\newcommand{\SISSA}{SISSA - International School for Advanced Studies, Via Bonomea 265, 34136 Trieste, Italy}
\newcommand{\SLAC}{SLAC National Accelerator Laboratory, Menlo Park, CA 94025}
\newcommand{\SMU}{Southern Methodist University, Dallas, TX 75275}
\newcommand{\SNOLAB}{SNOLAB, Lively, ON P3Y 1N2, Canada}
\newcommand{\Stanford}{Department of Physics, Stanford University, Stanford, CA 94305}
\newcommand{\StonyBrook}{Stony Brook University, Stony Brook, NY 11794}
\newcommand{\STSCI}{Space Telescope Science Institute, Baltimore, MD 21218}
\newcommand{\SUNYA}{University at Albany SUNY, Albany, NY 12222}
\newcommand{\SussexAstronomy}{Astronomy Centre, School of Mathematical and Physical Sciences, University of Sussex, Brighton BN1 9QH, United Kingdom}
\newcommand{\Syracuse}{Syracuse University, Syracuse, NY 13244}
\newcommand{\Tamu}{Texas A\&M University, College Station, TX 77843 }
\newcommand{\Techsource}{Techsource Incorporated, Los Alamos, NM 87544}
\newcommand{\TelAviv}{Tel-Aviv University,  69978 Tel-Aviv, Israel}
\newcommand{\Temple}{Temple University, Philadelphia, PA 19122}
\newcommand{\TIFR}{Tata Institute of Fundamental Research, Homi Bhabha Road, Mumbai 400005 India}
\newcommand{\Tsinghua}{Department of Astronomy, Tsinghua University, Beijing 100084, P R China}
\newcommand{\TUM}{Technical University of Munich,  80333 Munich, Germany}
\newcommand{\UA}{University of Alabama, Tuscaloosa, AL 35487}
\newcommand{\UAS}{Department of Astronomy/Steward Observatory, University of Arizona, Tucson, AZ  85721}
\newcommand{\UAM}{Universidad Aut\'onoma de Madrid, 28049, Madrid, Spain}
\newcommand{\UBC}{University of British Columbia, Vancouver, BC V6T 1Z1, Canada}
\newcommand{\UCB}{Department of Astronomy, University of California Berkeley, Berkeley, CA 94720, USA}
\newcommand{\UCBP}{Department of Physics, University of California Berkeley, Berkeley, CA 94720, USA}
\newcommand{\UCBSSL}{Space Sciences Laboratory, University of California Berkeley, Berkeley, CA 94720, USA}
\newcommand{\UCBGEN}{University of California Berkeley, Berkeley, CA 94720, USA}
\newcommand{\UCD}{University of California at Davis, Davis, CA 95616}
\newcommand{\UChicago}{University of Chicago, Chicago, IL 60637}
\newcommand{\UCI}{University of California, Irvine, CA 92697}
\newcommand{\UCLA}{University of California at Los Angeles, Los Angeles,  CA 90095}
\newcommand{\UCL}{University College London, WC1E 6BT London, United Kingdom}
\newcommand{\UCR}{University of California at Riverside, Riverside, CA 92521}
\newcommand{\UCSB}{University of California at Santa Barbara, Santa Barbara, CA 93106}
\newcommand{\UCSC}{University of California at Santa Cruz, Santa Cruz, CA 95064}
\newcommand{\UCSD}{University of California San Diego, La Jolla, CA 92093}
\newcommand{\UFL}{University of Florida, Gainesville, FL 32611}
\newcommand{\UFN}{Universit\`a Federico II di Napoli, 80125 Napoli, Italy}
\newcommand{\UGTO}{Divisi\'on de Ciencias e Ingenier\'ias, Universidad de Guanajuato, Le\'on 37150, M\'exico}
\newcommand{\UKY}{University of Kentucky, Lexington, KY 40506}
\newcommand{\UMD}{University of Maryland, College Park, MD 20742
\newcommand{\UMiami}{University of Miami, Coral Gables, FL 33124}}
\newcommand{\Umich}{Department of Physics, University of Michigan, 450 Church St, Ann Arbor, MI 48109}
\newcommand{\UmichAst}{Department of Astronomy, University of Michigan, 1085 S. University Ave, Ann Arbor, MI 48109}

\newcommand{\UMN}{University of Minnesota, Minneapolis, MN 55455}
\newcommand{\UnB}{Instituto de F\'{i}sica, Universidade de Bras\'{i}lia, 70919-970, Bras\'{i}lia, DF, Brazil}
\newcommand{\UNC}{University of North Carolina at Chapel Hill, Chapel Hill, NC 27599}
\newcommand{\UNH}{University of New Hampshire, Durham, NH 03824}
\newcommand{\UNIPD}{Dipartimento di Fisica e Astronomia ``G. Galilei'',Universit\`a degli Studi di Padova, via Marzolo 8, I-35131, Padova, Italy}
\newcommand{\UNM}{University of New Mexico, Albuquerque, NM 87131}
\newcommand{\UNV}{University of Nevada, Reno, NV 89557}
\newcommand{\UoM}{Jodrell Bank Center for Astrophysics, School of Physics and Astronomy, University of Manchester, Oxford Road, Manchester, M13 9PL, UK}
\newcommand{\UPenn}{Department of Physics and Astronomy, University of Pennsylvania, Philadelphia, Pennsylvania 19104, USA}
\newcommand{\UR}{Department of Physics and Astronomy, University of Rochester, 500 Joseph C. Wilson Boulevard, Rochester, NY 14627, USA}
\newcommand{\UrbanaC}{Department of Physics, University of Illinois at Urbana-Champaign, Urbana, Illinois 61801, USA}
\newcommand{\USC}{The University of South Carolina, Columbia, SC 29208}
\newcommand{\USD}{The University of South Dakota, Vermillion, SD 57069}
\newcommand{\UTD}{University of Texas at Dallas, Texas 75080}
\newcommand{\Utenn}{The University of Tennessee, Knoxville, TN 37996}
\newcommand{\Utah}{University of Utah, Department of Physics and Astronomy, 115 S 1400 E, Salt Lake City, UT 84112, USA}
\newcommand{\UVA}{University of Virginia, Charlottesville, VA 22903}
\newcommand{\Uvic}{University of Victoria, BC V8P 5C2 Victoria, Canada}
\newcommand{\UWaterloo}{Department of Physics and Astronomy, University of Waterloo, 200 University Ave W, Waterloo, ON N2L 3G1, Canada}
\newcommand{\UWMadison}{Department of Physics, University of Wisconsin - Madison, Madison, WI 53706}
\newcommand{\UW}{University of Washington, Seattle 98195}
\newcommand{\UWC}{Department of Physics \& Astronomy, University of the Western Cape, Cape Town 7535, South Africa}
\newcommand{\Vanderbilt}{Physics \& Astronomy Department, Vanderbilt University, PMB 401807, 2301 Vanderbilt Place, Nashville, TN 37235}
\newcommand{\VSI}{Van Swinderen Institute for Particle Physics and Gravity, University of Groningen, Nijenborgh 4, 9747~AG~Groningen, The~Netherlands}
\newcommand{\VT}{Virginia Tech, Blacksburg, VA 24061}
\newcommand{\VUU}{Virginia Union University, Richmond, Virginia, 23220}
\newcommand{\WCA}{Centre for Astrophysics, University of Waterloo, Waterloo, Ontario N2L 3G1, Canada}
\newcommand{\Weizmann}{Weizmann Institute of Science, 76100 Rehovot, Israel}
\newcommand{\Wellesley}{Wellesley College, Wellesley, MA 02481}
\newcommand{\wiscIce}{University of Wisconsin, Madison, WI 53706}
\newcommand{\WM}{College of William and Mary, Newport News, VA 23606}
\newcommand{\WUSL}{Washington University in St Louis, St. Louis, MO 63130}
\newcommand{\WVU}{CSEE, West Virginia University, Morgantown, WV 26505, USA}
\newcommand{\WVUGWAC}{Center for Gravitational Waves and Cosmology, West Virginia University, Morgantown, WV 26505, USA}
\newcommand{\Wyoming}{Department of Physics and Astronomy, University of Wyoming, Laramie, WY 82071, USA}
\newcommand{\Yale}{Department of Physics, Yale University, New Haven, CT 06520}
\newcommand{\Brandeis}{Department of Physics, Brandeis University, Waltham, MA 02453}
\newcommand{\NASAEinstein}{NASA Einstein Fellow}

\noindent
$^{1}$ \LBL \\
$^{2}$ \Yale \\
$^{3}$ \UR \\
$^{4}$ \Carnegie \\
$^{5}$ \NOAO \\
$^{6}$ \UCL \\
$^{7}$ \FNAL \\
$^{8}$ \Tsinghua \\
$^{9}$ \UAS \\
$^{10}$ \UCSD \\
$^{11}$ \Umich \\
$^{12}$ \UCBSSL \\
$^{13}$ \UWC \\
$^{14}$ \EPFL \\
$^{15}$ \CITA\\
$^{16}$ \PI \\
$^{17}$ \UCSC \\
$^{18}$ \IRFU \\
$^{19}$ \OSU \\
$^{20}$ \Wyoming \\
$^{21}$ \Pitt \\
$^{22}$ \UCBP \\
$^{23}$ \NASAEinstein \\
$^{24}$ \KIPAC \\
$^{25}$ \SLAC \\
$^{26}$ \UCB \\
$^{27}$ \OU \\
$^{28}$ \UFL \\
$^{29}$\UmichAst\\
$^{30}$ \Stanford \\
$^{31}$ \CfA \\
$^{32}$ \Siena \\
$^{33}$ \KSU \\


\medskip

 \begin{Abstract}
\noindent
In this white paper, we present an experimental road map for spectroscopic experiments beyond DESI.
DESI will be a transformative cosmological survey in the 2020s, mapping 40 million galaxies and quasars and capturing a significant fraction of the available linear modes up to $z=1.2$.
DESI-II will pilot observations of galaxies both at much higher densities and extending to higher redshifts.
A Stage-5 experiment would build out those high-density and high-redshift observations, mapping hundreds of millions of stars and galaxies in three dimensions, to address the problems of inflation, dark energy, light relativistic species, and dark matter.
These spectroscopic data will also complement the next generation of weak lensing, line intensity mapping and CMB experiments and allow them to reach their full potential.

\end{Abstract}

\section{Introduction}

Optical spectroscopy has been a cornerstone of modern cosmology experiments for the past 30 years, with multi-fiber spectroscopy greatly accelerating this reach. 
Spectroscopy has provided critical information in building our modern picture of structure formation and collisionless cold dark matter (CDM).
Spectroscopy provided the redshifts to the Type Ia supernovae in the discovery and further refinement of dark energy.
Multi-object spectroscopy from 2dF and SDSS produced the first maps of the universe with sufficient volume to measure the expansion rate with the Baryon Acoustic Oscillation (BAO) method.
These same redshift maps probe the evolution of redshift space distortions (RSD) over more than ten billion years, providing growth of structure constraints that complement BAO as a cosmological probe.  Beyond BAO and RSD, these maps can be combined with weak lensing and CMB maps to capture physics over a wider range of scales, thus offering accurate reconstruction of the power spectrum that encodes the initial conditions of the Universe produced by inflationary physics.

The next generation of spectroscopic experiments will improve upon these measures at $z<2$, and also dramatically extend our coverage of the high-redshift universe. Redshift-space maps at $z>2$ can
radically improve constraints on inflation and dark energy \cite{Ferraro:2019uce, Sailer:2021yzm, Ferraro:2022cmj, Linder:2021syd, Denissenya:2022ens}, while at the same time relaxing assumptions such as a power-law primordial power spectrum \cite{Slosar:2019gvt, dePutter:2014hza,Brando:2020yvo}. 

\begin{figure}[htb]
\includegraphics[width=16cm]{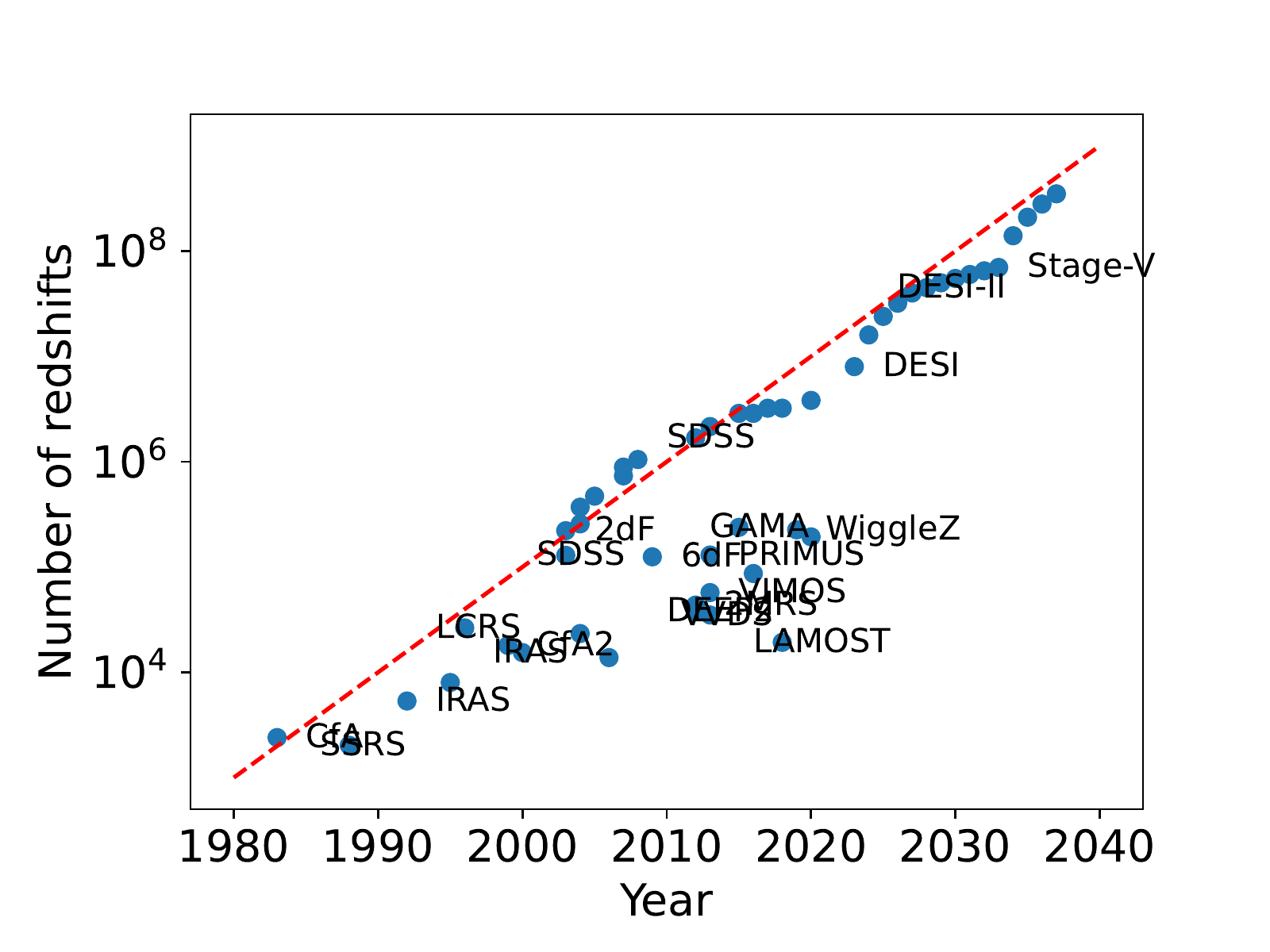}
\caption{Number of galaxy redshifts as a function of time for the largest cosmology surveys. The dotted line represents an increase of survey size by a factor of 10 every decade. Fielding a Stage-5 experiment in ten years maintains this pace into the 2030s, enabling the science cases discussed in this and other white papers.}
\label{fig:spectro_history}
\end{figure}

An upgrade to DESI, DESI-II, which we advocate as a first step in the future spectroscopic roadmap, will significantly extend the science reach of DESI, and serve as a pathfinder for larger Stage-5 spectroscopic survey. DESI-II will collect $\sim 40$ million redshifts at higher density and higher redshift than DESI. A Stage-5 survey will measure hundreds of millions of redshifts, opening up a new and uncharted discovery space. Figure \ref{fig:spectro_history} shows the history spectroscopic surveys, which have increased in size by an order of magnitude every decade so far. Fielding DESI-II and Stage-5 spectroscopy will maintain this pace and allow the transformative science described below. 

The execution of large-scale spectroscopic surveys necessitates imaging surveys with sufficient sky coverage, wavelength range, and depth to select appropriate targets.
SDSS provided much of the imaging for target selection for spectroscopic experiments from 2000 through 2020 \citep[e.g.][]{2002AJ....124.1810S,2002AJ....123.2945R,2009MNRAS.392...19C,2012ApJS..199....3R,2015ApJS..221...27M,2016ApJS..224...34P,2017MNRAS.471.3955R}. More recently, DESI targeting \citep{desitarget,dtelg,dtlrg,dtqso,dtbgs} --- which required deeper, better-calibrated imaging than SDSS could provide --- relied on the Legacy Imaging Surveys \citep{Dey:2019} combined with astrometric and photometric information from Gaia \citep{2018A&A...616A...1G,2021A&A...649A...1G}. Ultimately, in the next decade, the Rubin Observatory's Legacy Survey of Space and Time (LSST) \citep[e.g.][]{2019ApJ...873..111I} will begin to enable hundreds of millions of tracers to be targeted for Stage-5 experiments. In the interim, HSC \citep[e.g.][]{2022PASJ...74..247A} is available to facilitate pilot surveys that reach LSST depth over a smaller area.

\section{Science Drivers}

\begin{figure}[htb]
\begin{center}
\includegraphics[width=14cm]{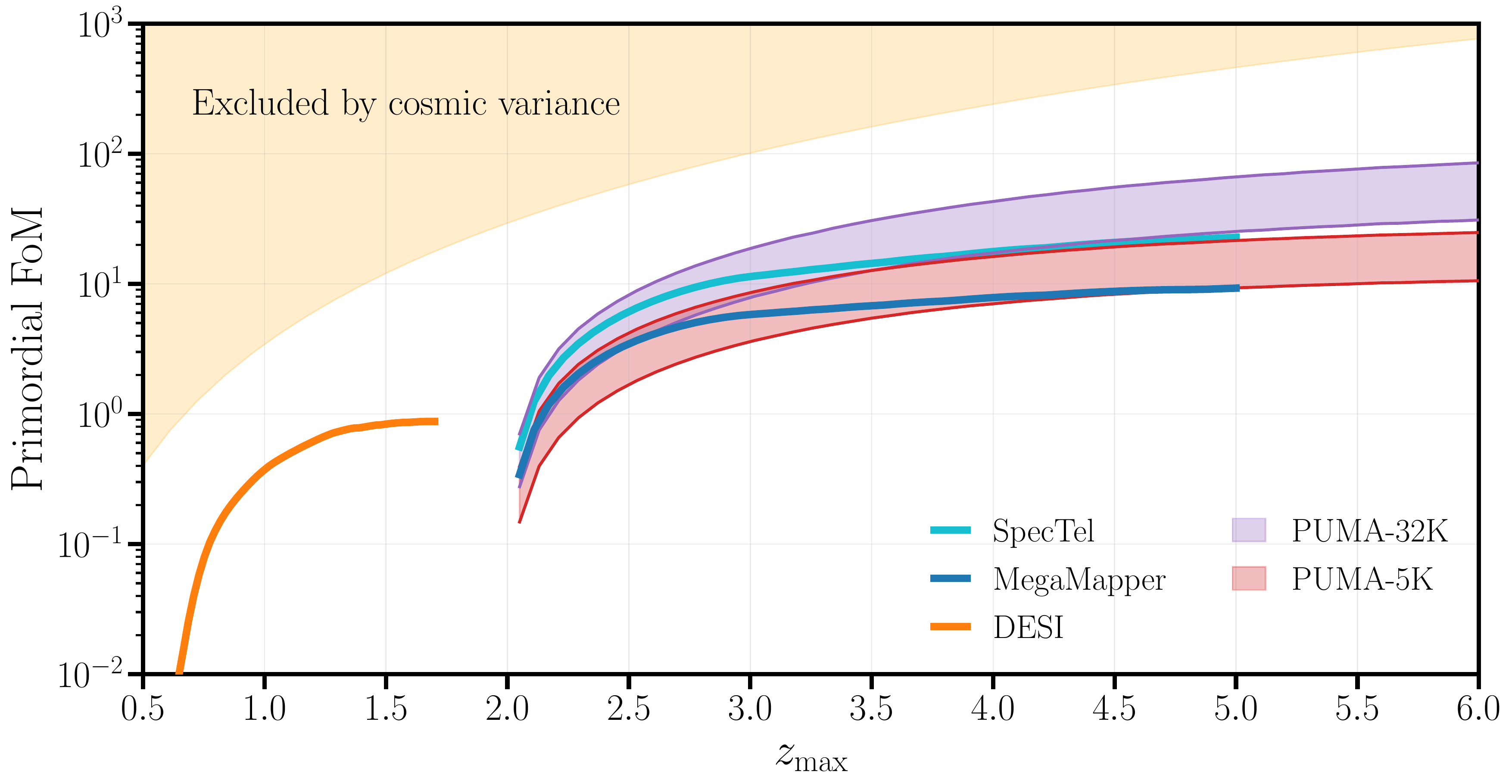}
\end{center}
\caption{``Figure of Merit'' FoM $\equiv 10^{-6}\,N_\text{modes}$, representing the effective number of ``linear'' modes observable as a function of $z_\text{max}$ for DESI, PUMA (-5K and -32K), and MegaMapper and SpecTel, two examples of Stage-5 spectroscopic surveys.  For DESI we include only the ELGs. For PUMA, we consider both optimistic and pessimistic foreground models, which are the boundaries of the shaded regions. The boundary of the shaded orange region is the cosmic variance limit for an all-sky survey, assuming $b(z)=1$. From \cite{Ferraro:2022cmj} (see \cite{Sailer:2021yzm} for details about the calculation).}
\label{fig:modes2}
\end{figure}

The mechanisms driving the accelerated expansion of the universe in its very first moments (inflation) and at late times (dark energy) represent some of the most important open problems in fundamental physics, and have been the subject of several of the Snowmass Community Science White Papers \citep[e.g.,][]{Achucarro22,Amin22,Blazek22,Dawson22, Ferraro:2022cmj}. At the same time, future surveys will allow us to measure neutrino masses to unprecedented precision, search for extra relativistic particles and greatly increase our understanding of dark matter and many astrophysical processes. In what follows, we summarize the science case, primarily based on previous work \cite{Sailer:2021yzm, Ferraro:2022cmj, Schlegel:2019eqc, Ferraro:2019uce}. We refer the reader to \cite{Sailer:2021yzm} for many of the details about the assumptions and forecasting methods, as well as a much more in-depth discussion of the science case, especially regarding the high-redshift sample. We start by noting that the constraining power on much of the fundamental physics described below is proportional to the number of linear or mildly non-linear modes that are correlated with the initial condition, which can therefore be taken as a ``Figure of Merit'' (FoM) when comparing current and future surveys. Figure \ref{fig:modes2} shows that the next generation of spectroscopic surveys can lead to an order of magnitude increase in the FoM, drastically improving our understanding of the universe.

\subsection{Inflation and Primordial Physics}

DESI will map the Universe to $z \sim 1.6$ at high fidelity. The large volume available at $1.6 \lesssim z \lesssim 5$, together with the existence of galaxy samples that can be targeted spectroscopically with existing technology, will allow the surveys proposed here to constrain primordial physics to unprecedented precision, beyond the CMB cosmic variance limit \cite{Ferraro:2022cmj} 
as spectroscopy unleashes the third dimension of spatial volume. 

Primordial non-Gaussianity \cite{Meerburg:2019qqi,Achucarro22} is one of the most powerful tools to study inflation and the primordial universe; the precise form of non-Gaussianity encodes the masses, spins and interactions of particles present during inflation. Due to the non-linear nature of gravity, inflation predicts a minimum amount of non-Gaussianity two orders of magnitude below current limits~\cite{Cabass:2016cgp}, with many models predicting much larger signals.  These deviations from Gaussianity leave specific imprints on the galaxy power spectrum and bispectrum \cite{Meerburg:2019qqi, Alvarez:2014vva, Dalal:2007cu, Ferraro:2014jba} that enable the measurement of particle physics of the inflationary era through these high-fidelity maps. It is through these surveys that ``cosmological collider physics"~\cite{Arkani-Hamed:2015bza} can probe scales inaccessible to Earth-based colliders. 

A particularly sharp boundary between inflationary models is set by the single-field consistency conditions~\cite{Maldacena:2002vr,Creminelli:2004yq}. They constrain the soft momentum limits of any $N$-point correlation functions if the inflaton is the only light degree of freedom.  Any deviation from these conditions requires new degrees of freedom, i.e. multi-field inflation. A specific example of this type is ``local'' non-Gaussianity, which requires an additional light scalar field. Without fine-tuning, multi-field inflationary models produce $\sigma(f_{NL}^{\rm local}) \gtrsim 1$, such that the full (untuned) parameter space can be explored with an order-of-magnitude improvement over current constraints \cite{Akrami:2019izv}. The presence of additional massive fields also gives rise to observable signatures that lie between local and equilateral type non-Gaussianity~\cite{Chen:2009zp,Meerburg:2016zdz,Gleyzes:2016tdh}.

The high redshift sample described in this paper can cross the important theoretical threshold of $\sigma(f_{NL}^{\rm local}) \lesssim 1$ from measurement of the galaxy power spectrum on large scales,\footnote{It has been pointed out \cite{Barreira:2022sey} that scale-dependent bias in LSS surveys provides a direct measurement of $b_\phi f_{NL}^{\rm local}$, where $b_\phi$ is the response of galaxy number density to local $f_{NL}$. In this work we fix $b_\phi$ to its fiducial peak-background split value $b_\phi \approx 2\delta_c (b -1)$, where $\delta_c$ is the critical density, and $b$ is the linear galaxy bias \cite{Dalal:2007cu}, but forecasts are easy to rescale for any chosen value of $b_\phi$.} surpassing the current CMB bounds by an order of magnitude \cite{Ferraro:2019uce, Sailer:2021yzm}. Potentially large improvements are possible when including the bispectrum, but we leave detailed forecasts to future work.
Improvements by a factor of two or larger over the current bounds are also expected for the equilateral and orthogonal shapes \cite{Ferraro:2019uce}.
Precise measurements of primordial non-Gaussianity will require precise control of systematics in the data and the analysis in addition to the sheer scale of the required surveys.

Deviations from scale-invariance of the initial conditions, particularly in the form of oscillations or features, are an additional opportunity to test inflation with future galaxy surveys. Inflationary models are famously sensitive to physics above the Planck scale and, in controlled examples, high-energy-scale physics can give rise to additional deviations from scale invariance~\cite{Slosar:2019gvt, Sailer:2021yzm}. A greatly-improved measurement of the spectrum of matter perturbations will significantly increase our understanding of the primordial fluctuations, and can be used to test for any type of oscillation or feature in the primordial power spectrum, offering another probe of the physical mechanisms behind inflation. Galaxy surveys already provide the best constraint on these models~\cite{Beutler:2019ojk} and deeper surveys will surpass even a cosmic-variance-limited CMB survey by up to an order of magnitude. Finally, the trispectrum or 4-Point Correlation Function has recently been shown to be sensitive to parity-violation in the initial conditions \citep{cahn,hou_odd,philcox}, with an intriguing possible detection of non-zero parity violation in BOSS. The large future spectroscopic samples proposed here will enable definitive follow-up of this result and, if confirmed, help distinguish the particular inflationary mechanism producing the parity-violation (e.g. axions, Chern-Simons gravity, etc.).


\subsection{Dark Energy}

Our understanding of dark energy \cite{Slosar:2019flp} will greatly benefit from measuring the expansion and growth of fluctuations throughout cosmic history. A lever arm that extends from low redshift (dark-energy-dominated era) to high redshifts (matter-dominated era) will tightly constrain large classes of theories, and test possible modifications to General Relativity. Moreover, precision measurements of the matter power spectrum will provide tight constraints on early dark energy (EDE) models, providing a few percent measurement of the energy content all the way to $z$$\sim$$10^5$ \cite{Sailer:2021yzm, Ferraro:2022cmj,Linder:2021syd}.

Using a combination of Redshift-Space Distortions (RSD) and BAO, we can measure the fraction of dark energy to better than 2\% up to $z \approx 5$ using the high-$z$ sample (see Fig. 5 in \cite{Ferraro:2022cmj} and \S4.8 of \cite{Sailer:2021yzm}). Other notable improvements include a factor of two better determination of the spatial curvature (compared to DESI + Planck), and a factor of $\gtrsim 2.5$ improvement in the dark energy figure of merit (Tab. 4 in \cite{Ferraro:2019uce}). EDE reveals new degrees of freedom from the time of radiation-matter equality to the late universe, and can offset the Hubble constant derived from CMB observations \cite{Abdalla:2022yfr}. 
A Stage-5 experiment will be able to constrain the fraction of EDE to better than $2\%$ all the way to $z \sim 10^5$ \cite{Sailer:2021yzm}, when the Universe was only a few years old, strongly testing this hypothesis, and more generally providing percent-level expansion constraints throughout cosmic history.

\subsection{Probing gravity and the amplitude of structure}

Modifications to General Relativity have been proposed as an alternative explanation for the observed accelerated expansion of the Universe at low redshift \cite{Jain10,Slosar19c}.  The phenomenology of such models can be very rich, but typically can lead to a decoupling between the growth of fluctuations and the standard predictions given the energy content determined by the cosmological parameters.  Measurements of the growth factor and its scale dependence through RSD over a wide redshift range is therefore a powerful probe of modified gravity \cite{Denissenya:2022ens}. Higher-order statistics such as the 3-Point Correlation Function, bispectrum, as well as marked 2-point statistics, can further tighten the constraints offered by a Stage-5 spectroscopic effort, as has been explored already in detail for DESI \citep{desi_mg_wp}.

Further tests are possible when combining 3D clustering measurements with lensing measurements (either galaxy or CMB lensing). A common ingredient of many modified gravity models is that the responses of the Newtonian potential $\Psi$ and spatial (Weyl) potential $\Phi$ to matter-energy (which are equal in GR) can differ: by comparing lensing by mass (sensitive to a combination of $\Psi$ and $\Phi$), to the motion of galaxies through RSD (sensitive to $\Psi$ only), one can look for deviations from the GR predictions. A Stage-5 survey, combined with CMB-S4, is sensitive to a few percent-level deviations up to $z \sim 5$ \cite{Sailer:2021yzm}.

\subsection{Neutrinos and other light relativistic species}

The presence of extra relativistic species at early times changes the radiation density (and hence affects the damping of the power spectrum on small scales), and also shifts the position of the BAO wiggles. Therefore the high-precision measurements of the power spectrum by future high-redshift surveys are an ideal way to detect the presence of extra light particles, with broad implications for fundamental physics \cite{Alexander16,Green19,Dvorkin:2022jyg}. The Standard Model of particle physics predicts $N_{\rm eff} = 3.044$ \cite{Bennett:2020zkv}, and parameterizing the number of extra species by $\Delta N_{\rm eff}$ \cite{CMBS4} lower limits exist on the change $\Delta N_{\rm eff}$ for broad classes of particles: $\Delta N_{\rm eff} > $ 0.027 for a single scalar, 0.047 for a Weyl fermion, and 0.054 for vector boson, even if thermal decoupling happens earlier than the rest of the Standard Model.\footnote{And a larger contribution if decoupling happens later than some of the Standard Model phase transitions.} As discussed in \cite{Ferraro:2022cmj, Sailer:2021yzm}, a Stage-5 high-$z$ survey can constrain $\sigma(N_{\rm eff})\approx 0.03$, and 0.02 when combined with proposed CMB experiments, reaching a sensitivity comparable to the smallest allowed value of $\Delta N_{\rm eff} = 0.027$ for a single particle. This sensitivity would exclude the thermalization of any particle with spin at $>$95\% confidence, which produces $\Delta N_{\rm eff} \geq 0.047$. If the universe was reheated above the QCD phase transition ($T\gtrsim 1$ GeV), such a measurement would also place the most stringent experiment constraints on a single axion coupled to any heavy fermion~\cite{Green:2021hjh,Dvorkin:2022jyg}.

Massive neutrinos leave a number of imprints on the CMB and LSS, allowing their total mass to be measured through cosmological observations. The most important effect is a suppression, proportional to the neutrino mass, of small scale power below the neutrino free-streaming scale. A measurement of this suppression can constrain the overall sum of neutrino masses to $\sigma(M_\nu) \approx 20$ meV or a $\approx 3\sigma$ ``detection''  in the case of the normal hierarchy, minimum mass  scenario (and more if the actual mass is larger). In $\Lambda$CDM+$M_\nu$, this measurement is limited by degeneracies, most significantly with the optical depth to reionization.  However, the suppression is redshift-dependent, allowing us to distinguish the effects of massive neutrinos from dynamical dark energy. Higher-order statistics such as the bispectrum are also sensitive to the way in which massive neutrinos change nonlinear mode coupling \citep{aviles, kama}. Probing a larger volume at high redshift allows a better determination of the large-scale power and will be more sensitive to a wide variety of scale dependent and/or redshift dependent effects from massive neutrinos~\cite{FrancoAbellan:2021hdb,Abazajian:2022ofy} and other new particles in the dark sector~\cite{Green:2021gdc,Dvorkin:2022jyg}.



\subsection{Fundamental Physics of Dark Matter}
\label{sec:darkmatter}

Determining the fundamental nature of dark matter represents one of the most significant unanswered problems in physics.
Astrophysical observations provide the only positive, empirical evidence of dark matter, and they provide a unique avenue for studying this mysterious component of the universe 
\cite{Buckley:2017ijx, Drlica-Wagner:2019xan, Bechtol:2019acd, Chakrabarti22, Bechtol:2022koa, Valluri_Chabanier_Irsic_Snowmass_DM_2022}.
Spectroscopic surveys of galaxies have contributed significantly to our understanding of dark matter and have been instrumental in developing the standard model of CDM.
However, in the expanding landscape of dark matter theories, many viable dark matter models can fit the observational data at the current level of precision, while predicting observable differences on small spatial scales beyond the reach of current experiments.
Models where dark matter interacts with itself or with Standard Model particles can predict appreciable changes in the shapes and central densities of dark matter halos.
Warm, fuzzy, and/or interacting dark matter models can suppress the abundance of small dark matter halos.
Furthermore, measurements of the distribution of dark matter in the Solar neighborhood and in other astrophysical systems are critical to interpret the results of direct and indirect detection experiments.

Wide-field, medium resolution spectroscopic surveys can provide unprecedented measurements of the radial velocities of stars in and around the Milky Way to greatly improve our understanding of the distribution and behavior of dark matter in the local Universe \cite{Li:2019nud,Valluri_Chabanier_Irsic_Snowmass_DM_2022}. 
In particular, spectroscopic surveys can:
(1) Measure the radial velocities of disk stars to constrain the local dark matter density and velocity distribution in the Solar neighborhood \citep{KuijkenGilmore89}. 
These measurements are critical for interpreting terrestrial dark matter experiments. 
(2) Measure the kinematics of tracer stars at large distances ($>75$ kpc) to determine the shape of the Milky Way's dark matter halo at large radii.  
Such measurements are sensitive enough to distinguish between collisionless and self-interacting dark matter models and are less sensitive to the baryonic physics of the disk \citep{2022MNRAS.tmp.2020V}.
(3) Combine radial velocity measurements of distant stellar tracers with proper-motion measurements from Rubin and Roman to decrease uncertainties on the total mass of the Milky Way halo. 
The total mass of the Milky Way halo is a key variable for interpreting measurements of the Milky Way subhalo mass function in the context of cosmological simulations.
(4) Measure the statistics of stellar streams and the dynamics of stars these streams to provide an exceptional probe of dark matter structure below the threshold of galaxy formation. 
The detection of dark matter halos without baryons would be a stunning confirmation of the CDM model, while a robust measurement of the absence of these halos would be strong evidence for new physics.

In addition to measurements within the Milky Way, wide-field spectroscopic surveys can also make detailed measurements of the shape and total mass of the dark matter halos of other Local Group galaxies. In addition, such surveys can effectively identify very low redshift galaxies and provide (1) a large sample of low redshift dwarf galaxies that can help constrain dark matter models and the galaxy halo connection and (2) a statistical sample of Milky Way-mass systems to put the Milky Way itself into essential cosmological context.

\subsection{Growth Constraints from Peculiar Velocities}

Peculiar velocity measurements in the $z\lesssim0.2$ universe 
have excellent complementarity with other large scale structure 
probes. They map the velocity field rather than the density 
field. By using individual object velocities rather than 
statistical clustering, they can overcome the cosmic variance
within the small volume at low redshift and thereby measure the growth rate, $f\sigma_8$, to well beyond the precision of galaxy clustering \cite{Kim:2019kls}. In combination with galaxy clustering measurements at high redshift, the resulting constraints on the structure growth index, $\gamma$, can be improved substantially (e.g., by $\sim2\times$ for DESI). Distances to sources come from 
calibrated distances, e.g.\ of Type Ia supernovae, 
gravitational wave sources, or galaxies through the fundamental 
plane or Tully-Fisher relations \cite{Kim:2019uxg,Kim:2019kls,Palmese:2020kxn}, 
while a multifiber spectroscopic survey obtains the 
required galaxy redshifts (and velocity dispersions). 

Spectroscopic standardization of Type~Ia supernovae provides the most accurate distances --- good to $\sim3\%$ per object \cite{snfactory21a, snfactory21b}. A spectroscopic instrument could provide spectra of nearby Type Ia supernovae when they are ``live'' by observing tiles in a target-of-opportunity mode. This would generate a few thousand Type~Ia supernova distances per year, and within a couple of years surpass in accuracy the velocity field that DESI will measure. As the surface density of ``live'' nearby supernovae is low ($<10^{-3}$\, deg$^{-2}$\,night$^{-1}$), only a single fiber would need to be transferred from a static target (galaxy or star) and placed on the live supernovae for each triggered tile. While individual spectra of live nearby Type~Ia supernovae are often obtained by other facilities, this approach would be unparalleled in its combination of depth and uniformity.

The operation of a full-sky, multi-object spectroscopic survey simultaneous with LSST will enable the next generation of peculiar velocity maps with only a small fraction of the fibers. Since the effects of cosmic acceleration and 
gravity beyond GR have their largest impact on growth at 
low redshift, including a peculiar velocity sample to the 
spectroscopic survey has substantial leverage---on thawing 
dark energy, dark energy with more extreme late time dynamics 
such as a phase transition, and testing theories of gravity 
in the both the scale-independent and scale-dependent 
regimes \cite{Kim:2019kls,Denissenya:2022ens}.

\section{Synergies with other cosmological probes}

\subsection{LSST and other imaging surveys}

\begin{figure}[htb]
\begin{center}
\includegraphics[width=12cm]{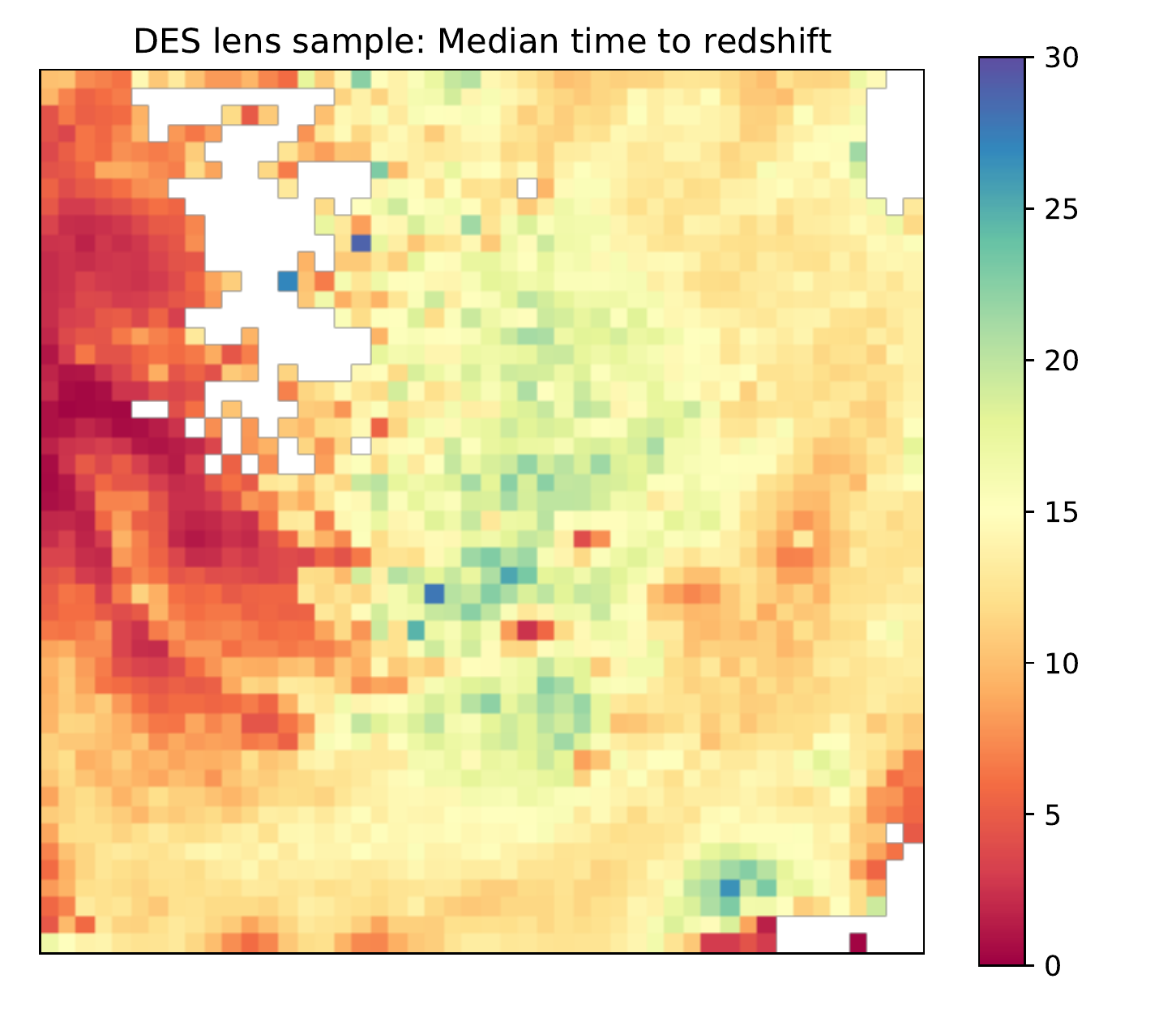}
\end{center}
\caption{Self-Organizing Map (SOM) trained on the DES Y3 MagLim lens sample via the algorithm of \cite{Sanchez:2020reg}, with color indicating the estimated median time in minutes for DESI to obtain a high-quality redshift for a galaxy in each cell at the sample's magnitude limit of $i=22.2$. The SOM is a non-linear 2D representation of the \textit{griz} color space with approximately equal number of galaxies in each cell. Cells with fewer than 10 spectra are shown in white. The SOM can be used to estimate observing time allocation as well as define photometric cuts where spectra are hard to obtain.}
\label{fig:dessom}
\end{figure}

A wide-field spectroscopic survey will greatly enhance the LSST science returns, as identified by several Astro 2020 science white papers \cite{Mandelbaum:2019zej, Newman:2019doi, Bechtol:2019acd}. Calibration of photometric redshifts is possible over the full range of LSST sources through cross-correlation techniques with the spectroscopic sample. A large overlap area will enable a reduction in the statistical errors to meet the stringent LSST requirements \cite{Mandelbaum:2019zej}. Furthermore, direct calibration of the redshift distributions of brighter samples of galaxies will be possible. For example, despite the implementation of an array of sophisticated techniques to estimate and calibrate the redshift distribution of the Dark Energy Survey (DES) lens and source galaxy samples, additional nuisance parameters were still necessary to absorb further systematic uncertainty.
However, $\sim80\%$ of the fiducial DES lens sample could have good redshifts obtained within 30 min.\ using DESI. Self-Organizing Maps (SOMs) can be used not only to estimate photometric redshift distributions as in \cite{DES:2020ebm}, but to estimate a wide range of predicted spectroscopic characteristics for galaxies as a function of their photometric properties, including time-to-redshift and redshift failure rates (Fig.~\ref{fig:dessom}). This in turn enables the precise tuning of photometric samples, e.g. to those objects for which redshifts can be easily obtained, and hence spectroscopic follow-up can directly and efficiently measure the redshift distribution.

While direct measurement of the faint LSST sources is likely infeasible, cross-correlation with a large-area spectroscopic sample will better constrain Intrinsic Alignment models.
Moreover, a combination of lensing amplitude provided by LSST, together with growth measurements through RSD can provide a powerful test of General Relativity on cosmological scales. Joint analysis of RSD and galaxy--galaxy lensing measurements will break degeneracies between galaxy bias and cosmological parameters, significantly mitigating cosmological uncertainty related to unknown galaxy-formation physics.
Finally, spectroscopy can provide redshifts for strong gravitational lenses and type Ia supernovae discovered in LSST, essential for their cosmological use.

\subsection{21cm and mm-wave Line Intensity Mapping}

Line Intensity Mapping (LIM) techniques, specifically 21cm and CO mapping, have significant potential as cosmological tools owing to the large number of accessible modes across a wide range of redshifts and physical scales. This measurements are challenging due to a number of instrumental and astrophysical issues, but we anticipate the technique will reach maturity over the next decade, proving itself a complementary and powerful standalone tool for cosmology.  Over this period, we further anticipate that wide-field spectroscopy will be {\it an essential complement} in order enable LIM to reach standalone potential.  In fact, the first cosmological results from CHIME have been obtained with stacking on eBOSS spectroscopic objects (LRGs, ELGs and quasars) \cite{CHIME:2022kvg}, and the first extragalactic analyses of the 21cm signal have been performed in cross-correlation with spectroscopic data \cite{2010Natur.466..463C, 2007MNRAS.376.1357L}. This is because the cross-correlation (or stacking) removes the large uncertainty due to foreground modeling and further highlighting the important role of spectroscopic catalogs for calibration and validation of 21cm experiments. Moreover, by accessing the largest scales, and therefore measuring the modes lost due to LIM foregrounds, they are highly complementary in terms of cosmological constraints. 


\subsection{CMB}

While future spectroscopic surveys provide compelling science reach on their own, it is also worth noting that the noise on CMB lensing maps from future wide-field experiments such as the Simons Observatory (SO) \cite{SimonsObs}, CMB-S4 \cite{CMBS4}, or a potential CMB-S5, 
will be reduced by more than one order of magnitude over existing measurements.  Unlike cosmic shear, CMB lensing can be measured to very high redshift, providing us with access to the matter field without the need to model bias.  Unfortunately CMB lensing alone mostly provides information that is projected along the line-of-sight (and with a broad redshift kernel).  Conversely, the cross-correlation of CMB lensing with LSS in several redshift bins (CMB lensing tomography; \cite{Yu:2018tem, Yu:2021vce}), can break the degeneracy with galaxy bias inherent in LSS alone; thus the cross-correlation provides measurement of the amplitude of mass perturbations as a function of redshift that cannot be obtained directly from either alone \cite{Modi:2017wds}.  This leads to tighter constraints on neutrino masses and dark energy, while mitigating some of the possible systematics \cite{Yu:2018tem,Wilson:2019brt,CMBS4}. Moreover, comparing the motion of non-relativistic matter through redshift-space distortions to the deflection of CMB photons will put some of the most informative bounds on theories of modified gravity \cite{Jain10,Chen:2022jzq}. 

At the same time, the cross-correlation of LSS with CMB lensing can potentially improve the robustness of constraints relying on the ultra-large scales, such as measurements of local non-Gaussianity \cite{Schmittfull:2017ffw}. In addition to CMB lensing, other secondary anisotropies\footnote{That is, fluctuations caused by the interaction of CMB photons with matter along the line of sight.} of the CMB correlate CMB and LSS maps, revealing the properties of the gas in the cosmic web (through the thermal and kinematic Sunyaev-Zel'dovich effects, tSZ and kSZ, and providing a tool to measure large-scale halo velocities through the kSZ \cite{Smith:2018bpn} and ``moving lens'' \cite{1983Natur.302..315B, Hotinli:2018yyc} effects. These velocity measurements can provide access to the very largest scales, often the most affected by primordial physics. In combination with LSS, they can be used to reduce cosmic variance, potentially providing an independent measurement of local primordial non-Gaussianity with $\sigma(f_{\rm NL}^{\rm loc}) \lesssim 1$ \cite{Munchmeyer:2018eey}.

\subsection{Gravitational waves}
\begin{figure}[htb]
\begin{center}
\includegraphics[width=10cm]{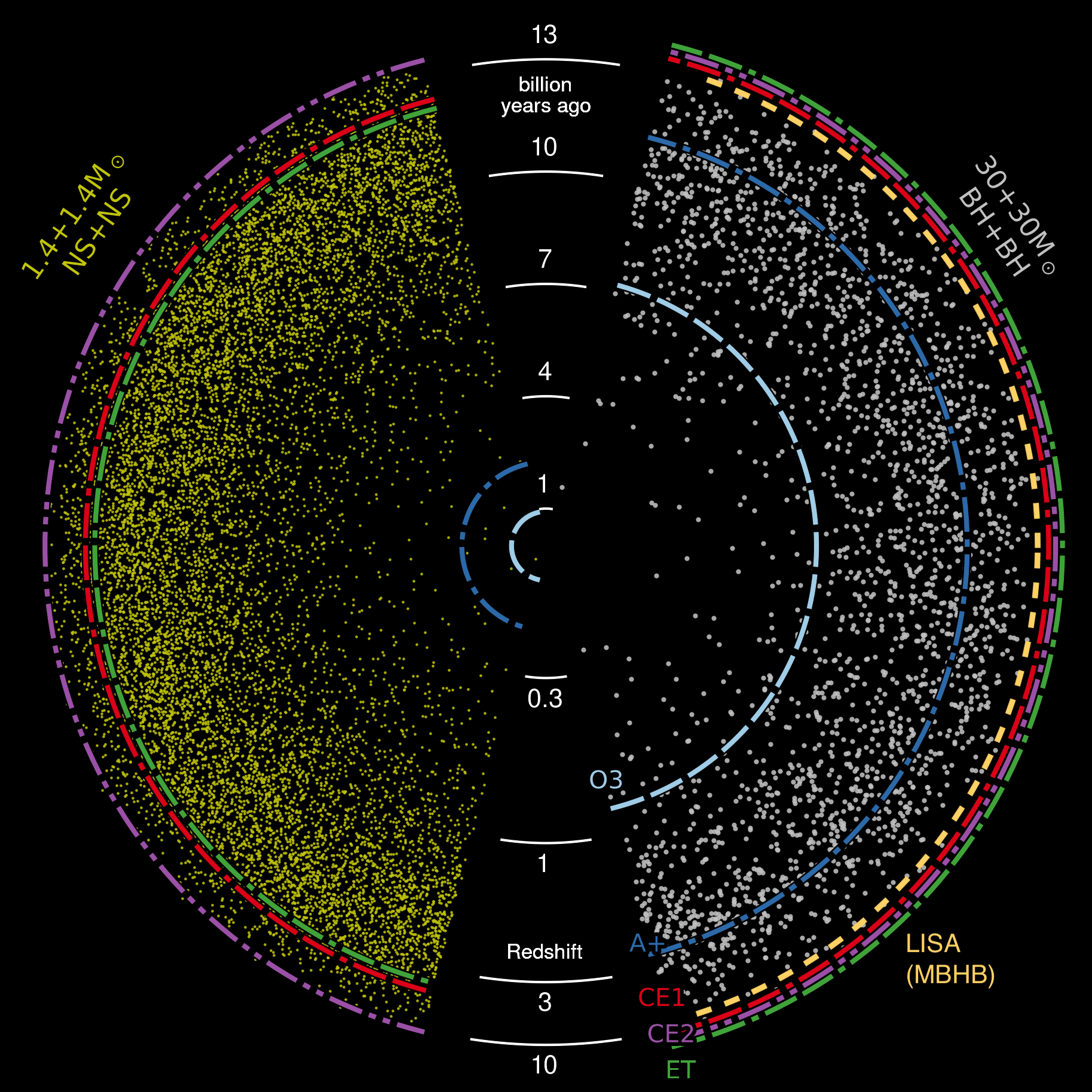}
\end{center}
\caption{Redshift reach of past and upcoming GW campaigns and experiments in the 2020s and 2030s. These include the LIGO/Virgo/KAGRA third observing run (O3; 2019-2020), the LIGO A+ upgrade (late 2020s), next generation ground-based detectors CE (including phase 1 and phase 2) and ET, and LISA. The left-hand side shows the expected distribution of BNS mergers, and right-hand side shows that of 30 $M_\odot$ black hole pairs. These mergers are expected to follow the cosmic star formation rate density, hence peaking around $z\sim 2$ (or less depending on the time delay between binary formation and merger). Hence a dense galaxy catalog out to $z\sim 1$, extending out to $z\sim 2$ even with smaller number density, would be an ideal complement to the upcoming GW detections for follow-up campaigns and cross-correlations. Note that the redshift reach of LISA is for massive black hole binaries (not the stellar mass black holes shown for the ground-based detectors). Credits: Evans \& Vitale, modified with permission.}
\label{fig:GWhorizon}
\end{figure}

 Between the 2020s and 2030s, hundreds of thousands of gravitational wave (GW) inspirals and mergers from binary systems will be detected across the gravitational wave spectrum. These will include hundreds of new binary neutron star (BNS), binary black hole (BBH), and neutron star-black hole (NSBH) mergers by the current Laser Interferometer Gravitational-Wave Observatory (LIGO \cite{ligo92,aligo}), Virgo \cite{Acernese_2014}, and KAGRA \cite{kagra} detectors, all of the merging BNS and BBH in the Universe with next generation gravitational wave detectors \cite{SnowmassGWdetectors} (such as the Cosmic Explorer \cite{cosmicexplorer}), intermediate mass and supermassive BBH mergers by the Laser Interferometer Space Antenna (LISA \cite{lisa}) and Pulsar Timing Array \cite{ipta}, amongst other sources of gravitational waves. These detections will represent an extraordinary tool to probe fundamental physics and beyond the Standard Model physics in some of the most extreme environments \cite{BertiWP}.
 In particular, an enormous range of scientific analyses (e.g. \cite{LIGOScientific:2017ync}) is enabled when the electromagnetic counterpart to these GW sources can be identified, ranging from cosmological parameters inference, to tests of gravity, to the behaviour of nuclear matter and the progenitors of binaries, analyses that would otherwise be impossible or significantly less precise without the identification of the source or at least the host galaxy. Future spectroscopic surveys can play a unique role towards enabling the identification of electromagnetic counterparts and towards cosmological analyses \cite{Palmese_Astro2020,CF6_transients} through the following:\\
 \vspace{0.2pt}
{\bf A complete galaxy catalog}: a Stage-5 spectroscopic experiment will provide redshifts for galaxies accounting for a significant fraction of the stellar mass at $z<1$ (where the GW events will also provide better localizations compared to higher redshifts), allowing a refinement of the possible host galaxies of candidate electromagnetic counterparts identified through imaging 
 to a smaller number with redshifts consistent with the GW distances. The availability of a complete galaxy catalog also enables deep, targeted searches of counterparts within specific galaxies, as wide field ground based searches will become increasingly more challenging as the horizon of GW detectors improves. In some cases, for extremely well localized events or strongly lensed GW events, a complete redshift catalog may help us to identify the host galaxy even without an electromagnetic counterpart. When the host galaxy redshift is known, the GW events can be used as ``standard sirens'' \cite{schutz}, to derive measurements of the expansion rate of the Universe and constraints on modified gravity models.\\
 \vspace{0.2pt}
{\bf Real-time follow-up}: a Stage-5 spectroscopic platform would also provide the capability of a multiplexed triggered follow-up for thousands of galaxies and candidate electromagnetic counterparts within a GW localization region. This would enable rapid characterization of the candidate counterparts found by imaging surveys, which is particularly valuable for the identification of the rapidly fading ``kilonova'' \cite{Kasen_2017} transients expected to follow BNS mergers. \\
\vspace{0.2pt}
 {\bf Cross-correlations}: a unique contribution of Stage-5 spectroscopic experiments to multi-messenger cosmology will be that of enabling cross-correlations between the galaxy and the GW source density \cite{2017MNRAS.469..656R}. Even in the absence of electromagnetic counterparts, cross-correlations will not only provide measurements of cosmological parameters \cite{Raccanelli_2017}, but they will also allow us to probe the origin of BBH \cite{Raccanelli_2016}. Amongst the various applications, such analyses will clarify whether some BBH could be of primordial origin \cite{scelfo} and constitute part of the dark matter content of the Universe.\\
 \vspace{0.2pt}
 {\bf Discovering the Electromagnetic counterparts of LISA sources}: repeated spectroscopic observations of large samples of galaxies and stars are needed to identify and characterize a sample of sources that will be detectable with LISA in the late 2030s. First, catalogs of binary stars inspiraling in the LISA band are needed to create a sample of Galactic ``verification binaries'' \cite{Stroeer_2006} that can be used for studies of compact objects and calibration. Second, identification of samples of massive black hole binaries in the form of e.g. dual AGNs through the identification of Doppler shift in AGN lines will inform future follow-up observations and enable sub-threshold or targeted searches in the PTA and LISA signal.





\subsection{Discovery space}

Spectroscopic observations of hundreds of millions of objects furnish a discovery space for rare astrophysical phenomena, much of which will be useful for studying dark matter, dark energy, reionization, and cosmology. For example, SDSS provided discoveries of changing look quasars \citep{2015ApJ...800..144L}; dozens to hundreds of strong gravitational lenses \citep[e.g.][]{2012AJ....143..119I, Shu2017, 2021MNRAS.502.4617T}, some with novel configurations \citep{2003Natur.426..810I, Gavazzi2008}; dozens of high-redshift quasars \citep[e.g.][]{2016ApJ...833..222J}, some with Gunn-Peterson troughs \citep[e.g.][]{2001AJ....122.2850B}; thousands of proximate quasar pairs with which to probe clustering in the IGM \citep[e.g.][]{2018ApJS..236...44F}; UV-bright quasars to probe helium reionization \citep[e.g.][]{2011ApJ...728...23W}; massive overdensities at high redshift \citep[e.g.][]{2017ApJ...839..131C}; Damped Lyman-alpha systems \citep[e.g.][]{2018MNRAS.473.3019P}; vanishing Broad Absorption Line quasars \citep{2012ApJ...757..114F}; and candidate recoiling, coalescing or binary supermassive black holes \citep[e.g.][]{2008ApJ...678L..81K, 2009Natur.458...53B, 2009ApJ...697..288B, 2009MNRAS.398L..73D}. SDSS also provided spectroscopic discoveries of scores of transients near galactic centers \citep[e.g.][]{2013MNRAS.430.1746G}. Beyond the potential identification of new kinds of transients, these measurements constrain the volumetric rates, physical origins, and host galaxy properties of standardizable transients used to measure cosmic expansion and the growth of structure.

In the LSST era, combined imaging and spectroscopy of these phenomena will create novel and powerful tools for cosmology. Strong lensing systems provide an illustrative case, with imaging data measuring their unique spatially and temporally resolved phenomenology and spectroscopy providing the component redshifts that are needed to convert angular measurements into mass measurements. Imaging and spectroscopic discovery of strong lenses are also highly complementary in the angular scales and flux contrast ratios to which they are sensitive. Lensed transients such as AGN flares and supernovae probe cosmic expansion using time delay cosmography (see e.g. \cite{2022arXiv220713489L} and references therein). LSST is projected to observe $\mathcal{O}(100)$ strongly lensed supernovae per year, and more than half of those will occur in systems with small Einstein radii ($<1''$ separation) \cite{2019ApJS..243....6G}, making them difficult to identify in imaging data. High-statistics catalogs of spectroscopically discovered strong lensing systems will significantly increase the number of time-delay observations possible with LSST \cite{2021ApJ...910...65B}. Strong galaxy-galaxy lenses observed with both imaging and spectroscopy can also be used to constrain the dark matter halo mass function on subgalactic scales through the effect of surface-brightness perturbations \citep[e.g.][]{Vegetti2010, Vegetti2014}.

\section{Targets}

\subsection{Overview of target samples}

Ground-based multi-fiber spectroscopy offers the opportunity to efficiently observe cosmologically useful samples of objects by selectively targeting only those samples.  DESI is currently employing this strategy to target a sample of galaxies and quasars optimized for the study of dark energy.  The ground-based PFS and 4MOST spectrographs will do the same (on a smaller scale) when they begin operations. The space-based Euclid and Roman satellites will observe sets of intermediate-redshift galaxies limited to strong emission line galaxies.  Although the combination of surveys from these instruments will be transformative, mapping tens of millions of galaxies, they will only scratch the surface of the total potential information. 

Figure \ref{fig:modes2} quantifies this in terms of perturbative regime modes measured, considering bias and number density. This quantity can be taken as a ``Figure of Merit'' for comparing current and future surveys, since many of the constraints on the fundamental physics highlighted in this paper scale proportional to this quantity. 

The spectroscopic road map presented here builds out galaxy samples to both higher redshifts and higher densities to address the frontier of cosmology through the 2030s.

Notional samples of stars, galaxies and quasars are described below.  The suitability of these samples for successful spectroscopic observation have either already been demonstrated, or will be demonstrated in DESI pilot observations.
The flexibility of multi-fiber spectrographs means that the existing sample selections can be optimized in the future to the targets and redshift ranges of most interest.

\subsection{Details of target samples}

\subsubsection{Targets: Low-redshift galaxies}
\label{sec:targets-lowz}

Low-redshift ($z<1$) galaxies represent the easiest galaxies for which to efficiently measure redshifts.
``Bright'' in this context refers to those galaxies whose surface brightness is only marginally fainter than the dark night sky at optical wavelengths.
DESI is already observing a subset of these galaxies, split between a Bright Galaxy Sample (BGS) and a luminous Red Galaxy Sample (LRGS).  These samples have been selected using flux and color cuts to deliver galaxy samples at a density optimized for the DESI science case.
Removing color cuts yields samples of 6600 galaxies per sq. deg. to a z-band fiber magnitude limit of $z_{fiber} < 21.6$, and 14,000 galaxies per sq. deg. to a limit of $z_{fiber} < 22.4$.

These bright galaxy samples can be targeted from any of the modern, broad-band imaging surveys.  For DESI-II, these would be targeted from the Legacy Imaging Surveys.  For Stage-5, these could be targeted from either future data releases of the Legacy Imaging Surveys or from LSST.

Successful redshifts for these galaxies have been demonstrated in 1000 sec (bright sample) and 4000 sec (faint sample) exposures using the DESI instrument.  The observation times for Stage-5 are assumed to be shorter by a factor of 2.5.  Note that there is minimal contamination of this sample by stars.

An additional class of very low redshift galaxies, focused on $z<0.1$, can inform dark matter physics through measurements of dwarf galaxies, provide a large sample of peculiar velocities and a dense map of the very local Universe, and provide host galaxies for gravitational wave and transient sources, including those identified by LSST. Pilot programs with DESI have indicated that a complete magnitude-limited sample of $z<0.03$ galaxies can be obtained using photometry 
alone, with 
800 objects per deg$^2$ at $19 < r < 22$ \citep{desilowz}.

\subsubsection{Targets: Mid-redshift [O~II] emitters}
\label{sec:targets-midz}

Star-forming galaxies at redshifts $\sim 0.5-2$ are abundant and suitable targets for redshift surveys owing to their bright emission lines.
DESI is making use of the O~II 3727\AA\ emission line to measure redshifts for 20 million of these galaxies spanning redshifts $0.6 < z < 1.6$.
PFS should deliver a smaller sample of galaxies to higher redshift, with 4.5M emission line galaxies spanning redshifts $0.6 < z < 2.4$ within their 1500 sq deg. footprint \cite{2014PASJ...66R...1T}.
The Euclid satellite will make use of the H$\alpha$ 6563 emission line to measure redshifts for $\sim 4000$ galaxies per sq deg. over 15,000 sq deg over the redshift range $0.9<z<1.8$ to a flux limit of $2\times 10^{-16}$ erg~s$^{-1}$~cm$^{-2}$ \cite{2018MNRAS.474..177M}.
The Roman satellite will make use of the H$\alpha$ 6563 emission line to measure redshifts for $\sim 10,000$ galaxies per sq deg. over 2,000 sq deg over the redshift range $1<z<2$ to a flux limit of $1\times 10^{-16}$ erg~s$^{-1}$~cm$^{-2}$ \cite{2018MNRAS.474..177M}.
There are large uncertainties in the projections for these satellite programs, and the resulting samples will have non-negligible contamination rates from lower-redshift galaxies and complicated spatial completeness owing to blending of objects in slit-less grism spectroscopy\cite{2020ApJ...897...98B}.

The opportunity exists to extend the sample of DESI O~II emitters with the addition of a channel to the DESI spectrographs.  A 4th channel extending the wavelength coverage to 1.2 microns would make it possible to measure redshifts for samples to redshifts $z=2.2$.  These could be observed to luminosities and line fluxes fainter than the Euclid and Roman surveys, and would avoid the complex spatial completeness of those surveys.

Targeting ELGs in the redshift range $1.4 < z < 2.2$ requires deep, broad-band photometry as will be available from LSST.
Such photometry enables a selection with low contamination using simple cuts. Typically, a cut in the $g$-band controls the desired target density and favors strong [O~II] emitters -- hence ensuring a high redshift success rate.  Color cuts in the $gri$- or $grz$-diagram control the redshift range. For instance, \cite{2014PASJ...66R...1T} present a simple $g<24.2$ mag and $gri$-selection based on the HSC/Wide survey, optimized for the $0.6<z<2.4$ range.
We include in Table \ref{tab_sample_stage5} for Stage-5 spectroscopy a slightly different sample at $g<24.1$ mag, which yields 1000 galaxies per deg$^{-2}$ spanning the redshift range $1.4<z<2.2$.
Such a selection is possible from the data presented in \cite{2022ApJS..258...11W}, based upon HSC deep photometry in the COSMOS field degraded to the 10-year depth of LSST.

\subsubsection{Targets: Lyman-break galaxies (LBG)}
\label{sec:targets-lbg}

Dropout-selected or Lyman Break Galaxies (LBGs) have been a traditional sample of galaxies for mapping the universe at redshifts $z>2$ \cite{Giavalisco:2002si,Shapley:2011tq,Wilson:2019brt}.  Dropout selection on magnitude limited surveys naturally selects massive, actively star-forming galaxies that comprise the majority population over the redshift range of interest.  Except at the bright end, the fraction of obscured or reddened galaxies which are missed by the selection is small, though the completeness does vary with redshift \cite{Ly:2009jg,Wilson:2019brt}. The name is indicative of the spectroscopic feature of a break in the spectrum due to absorption from the hydrogen atom Lyman series from neutral hydrogen in stellar atmospheres, interstellar photoelectric absorption and Lyman-series blanketing along the line of sight \cite{Madau:1995js}.  That feature both makes it possible to select these galaxies at $z>2$ from sufficiently-deep broad-band imaging, and to measure redshifts with instruments such as DESI.  Due to the lack of other strong spectroscopic features, these do require relatively long exposure times to confidently measure redshifts \cite{Wilson:2019brt}.

For a notional sample of LBGs, we considered broad-band photometry would be available to 2-year LSST depth for DESI-II and to 10-year depth for Stage-5
Several LBG selection critera are presented in \cite{Wilson:2019brt} that are based upon cuts in color-magnitude space to provide samples at different redshifts.  We combine four of those selections for Stage-5 spectroscopy: a BX selection for $z\sim 2$ galaxies to a depth of $m_{\rm r}<24.5$ mag, a $u$-drop selection for $z\sim 3$ galaxies to a depth of $m_{\rm r}<24.5$ mag, a $g$-drop selection for $z\sim 4$ to a depth of $m_{\rm i}<25.0$ mag, and a $r$-drop selection for $z\sim 5$ to a depth of $m_{\rm z}<25.0$ mag.
For DESI-II, we include only the first two of these samples to shallower depths of $m_{\rm r}<24.0$ mag.
Note that these samples would dominate the observing time for DESI-II or Stage-5 if not further down-selected.


\subsubsection{Targets: Lyman-alpha emitters(LAEs)}
\label{sec:targets-lae}

The Universe has been kind in providing a large density of emission line galaxies at high redshifts \cite{1996MNRAS.283.1388M}.  These galaxies are not as massive as the typical LBG, but active star formation make these galaxies easy to redshift.  Dust in these galaxies reprocess young starlight into emission lines, with the strongest of those lines being the Lyman-alpha emission line that is visible in the optical at redshifts greater than 2.

Several projects are exploring these emission line galaxies at selected redshifts.  This includes the Konno et al.\ program at $z=2$ using the Subaru Telescope \cite{2016ApJ...823...20K}, the ODIN project at $z=$2.4, 3.1, and 4.5 using the Blanco Telescope\footnote{\url{https://legacy.noirlab.edu/noaoprop/abstract.mpl?2020B-0201?}}, and the GOLDRUSH and SILVERRUSH projects on the Subaru Telescope at even higher redshifts \cite{2018PASJ...70S..10O,2018PASJ...70S..13O}.
All of these projects employ deep, narrow-band imaging to identify Ly-$\alpha$ emission in a narrow redshift shell.  ODIN detects LAEs to a flux limit of $0.1-0.3 \times 10^{-16}$ erg~s$^{-1}$~cm$^{-2}$ that validates the Konno et.\ al luminosity function.  DESI has conducted pilot spectroscopic observations of those targets, demonstrating the efficiency of these selections and a high redshift success rate.

Targeting of these galaxies requires narrow band imaging to supplement available broad-band imaging.  Detection of light in a narrow band in excess to the broad-band flux is indicative of an emission line in that band.  The sheer volume of the universe at high redshift means the vast majority ($75\%$) of these selected objects are LAEs.
The DESI-II sample assumes that narrow band imaging is available from 4000--5000\AA, from which to select LAEs at redshifts $2.3<z<3.1$ to a Lyman-alpha flux limit of $1.0\times 10^{-16}$ erg~s$^{-1}$~cm$^{-2}$.  This is approximately the same flux limit as the O~II emitters in DESI, with reliable redshifts measured in 30 min. exposures.
The Stage-5 samples assumes that narrow band imaging is available in a larger 3800--5500\AA\  range, from which to select at redshifts $2.1<z<3.5$.  Extending the selection to a flux limit of $0.5\times 10^{-16}$ erg~s$^{-1}$~cm$^{-2}$ would require approximately the same 30 min. exposures on the larger aperture of Stage-5.

\subsubsection{Targets: Stars}
\label{sec:targets-stars}

The kinematics of stars has long been a complementary tool for understanding the bulk properties of dark matter.  Rotation curves, stellar velocity dispersions, disk kinematics, and stellar streams have all been demonstrated as methods for probing the smooth and structured properties of the dark matter distribution within galactic contexts.  The stellar halo of the Milky Way, owing to long dynamic times that preserve orbital memory and the modest contaminating baryonic effects, is recognized as a particularly fruitful site for experimental access to the dark sector.  While much progress has been made in the inner stellar halo and the disk of the Milky Way, thanks to surveys like SDSS and Gaia, to probe the dark-matter dominated regions of the Milky Way requires kinematic mapping to $i= 22.5$ AB mag in the halo.

The combination of LSST imaging and Stage-5 spectroscopy will be required to undertake the dark matter experiment as described in Section \ref{sec:darkmatter}.
Current and upcoming spectroscopic surveys are probing the Milky Way's halo (DESI, SDSS I-V, WEAVE, 4MOST, PFS) with a sampling of a few million stars with only a fraction being in the Milky Way halo.
Notionally, we propose targeting a large fraction of the Milky Way halo stars.
Estimates from TRILEGAL\footnote{ http://stev.oapd.inaf.it/cgi-bin/trilegal} indicate at the Galactic pole (l=0$^\circ$,b=90$^\circ$) there are approximately 4000 stars/deg$^2$ at $i\le22.5$~mag, with halo stars comprising $\approx$1366 stars/deg$^2$.
LSST will provide proper motion measurements to a precision of $<$0.5~mas/yr for these stars \cite{2012IAUS..282...33E}, while DESI and Stage-5 would provide velocity precision of $\sim 4$ km/s.
Tables  \ref{tab_sample_desi2} and \ref{tab_sample_stage5} list a notional survey of 7.5M of these stars as a DESI-II pilot survey, followed by 60M of these stars over a 10,000 sq. deg footprint as a portion of a Stage-5 spectroscopic program.
Once a more quantitative metric of the required density of stars is developed, the target numbers would be tuned to match those requirements.

\subsection{Observation strategy}

Multi-fiber spectrographs allow enormous flexibility in observing strategy.
Target samples can be divided into samples that are brighter, to be observed in worse observing conditions, and fainter samples for the better conditions.  Targets requiring longer exposures can be re-observed on multiple visits to the same location on the sky, and those data combined for measuring redshifts.
DESI makes use of these strategies, and it is assumed that DESI-II and Stage-5 would further optimize this approach.

DESI has demonstrated that it observes 700 ``effective hours'' per year, which corresponds to hours of dark, clear time with good (1.1 arcsec) seeing.
Bright, moon-up time is included in this number, weighted appropriately for the higher sky brightness.
DESI-II and Stage-5 should have this same number of effective hours per year with a survey program that makes dedicated use of the instrument.

DESI-II would operate with the same Mayall Telescope at Kitt Peak, either with the existing instrument or an upgraded instrument.
DESI utilizes $\sim$4300 fibers per exposure after accounting for fixed positioners used for sky-background subtraction and non-functional positioners.
With an 85\% utilization of these fibers for science targets,
this translates to 2.56M available fiber-hours per year.
Therefore, the entire DESI-II 62.8M would require approximately 25 years of observation.  A DESI Upgrade to 11,250 fibers would speed this program by a factor of 2.6.   The upgrade could occur mid-program, with the observing appropriately split pre- and post-upgrade.

The Stage-5 program should be achievable from a telescope location in either the northern or southern hemisphere.  Table \ref{tab_sample_stage5} outlines the required time if conducted with the MegaMapper instrument, with spectrographs identical to DESI in performance and a 6.5-m mirror collecting area.
The mid-redshift OII sample would only be accessible if the spectrographs were upgraded to include a 4th arm, extending the wavelength coverage from 0.98 to 1.2 microns.
Assuming 700 effective dark hours per year, an 85\% utilization of those fibers for science targets, and 26,100 fibers, one has 15.6M fiber-hours available per year.  Therefore the entirety of the tabulated program would take 18 years to perform using MegaMapper.  Obviously, judicious choices would be necessary to down-select these targets in optimizing a Stage-5 spectroscopic program.

\begin{table}[htb]
\caption{DESI-II samples}
\label{tab_sample_desi2}
\begin{tabular}{lrrrrrrr}\hline
 & Redshifts  & Density & Area & Number & Efficiency & Exptime & Fiber-hours \\
 &       & (deg$^{-2}$) & (deg$^2$) & & & (hr) &  \\
\hline
Stars i$<21.5$  &     0 &  500 & 10,000 & 5.0M & 1.0 & 1.5 & 7.5M  \\
Low-z bright  &     0--1.0 &  6600 & 1400 &  9.2M & 1.0 & 0.28 & 2.6M \\
Low-z faint   &     0--1.0 &  7900 & 1400 & 11.0M & 1.0 & 1.0 & 11.0M \\
LBGs m $<24.0$   &   2.0--4.5 &   1250 & 10,000 &  12.5M & 0.6 & 2.0 & 41M \\
LAEs f $> 1.0\times 10^{-16}$ &  2.3--3.1 &  800 &   1400 &  1.1M & 0.75 & 0.5 & 0.7M \\
\hline
Total         &            &       &          & 38.8M & & & 62.8M \\
\hline
\end{tabular}
\end{table}

\begin{table}[htb]
\caption{Stage-5 samples}
\label{tab_sample_stage5}
\begin{tabular}{lrrrrrrr}\hline
 & Redshifts  & Density & Area & Number & Efficiency & Exptime & Fiber-hours \\
 &       & (deg$^{-2}$) & (deg$^2$) & & & (hr) &  \\

\hline
Stars i$<22.0$  &          0 &  4000 & 10,000 & 40.0M & 1.0 & 1.5 & 60M  \\
Low-z bright  &     0--1.0 &  6600 & 18,000 & 119M & 1.0 & 0.1 & 11.9M \\
Low-z faint   &     0--1.0 &  7900 & 18,000 & 142M & 1.0 & 0.4 & 56.8M \\
Mid-z [O~II] &   1.4--2.2 &  1000 & 10,000 &  10M & 0.85 & 1.0 & 11.7M \\
LBGs m $<24.5,25.0$   &   2.0--6.0 &  4500 & 10,000 &  45M & 0.6 & 1.5 & 112.5M \\
LAEs f $> 0.5\times 10^{-16}$ &   2.1--3.5 &  5000 & 10,000 & 50M & 0.75 & 0.5 &   33.3M \\
\hline
Total         &            &       &          & 406M & & & 286M \\
\hline
\end{tabular}
\end{table}

\begin{figure}[htb]
\begin{center}
\includegraphics[width=13cm]{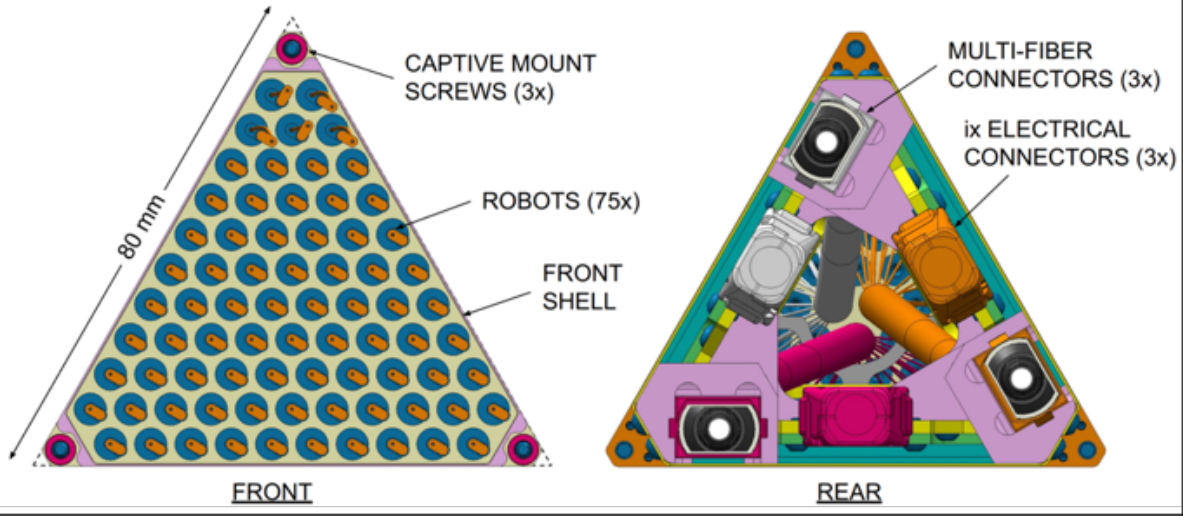}
\end{center}
\caption{Fiber robot raft design in development for a DESI upgrade or a Stage-5 instrument.
Each raft consists of 75 mechanical robots to position each fiber, with a center-to-center pitch of 6.1-mm.
}
\label{fig:rafts}
\end{figure}

\begin{figure}[htb]
\begin{center}
\includegraphics[width=12cm]{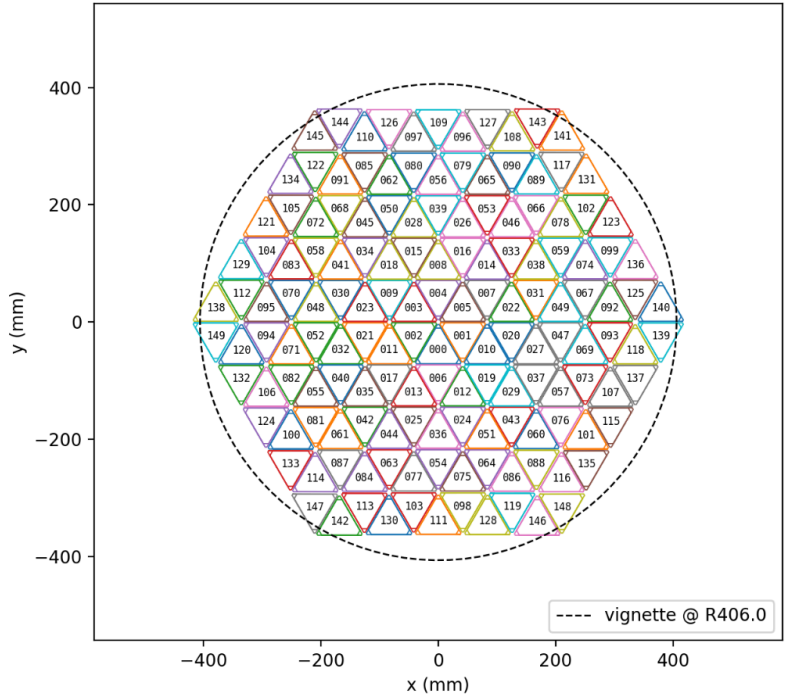}
\end{center}
\caption{Focal plane layout for 150 fiber robot rafts (with 11,250 robots) on the DESI focal plane.  These rafts would be re-usable for MegaMapper.}
\label{fig:desi2-focal}
\end{figure}

\begin{figure}[htb]
\begin{center}
\includegraphics[width=14cm]{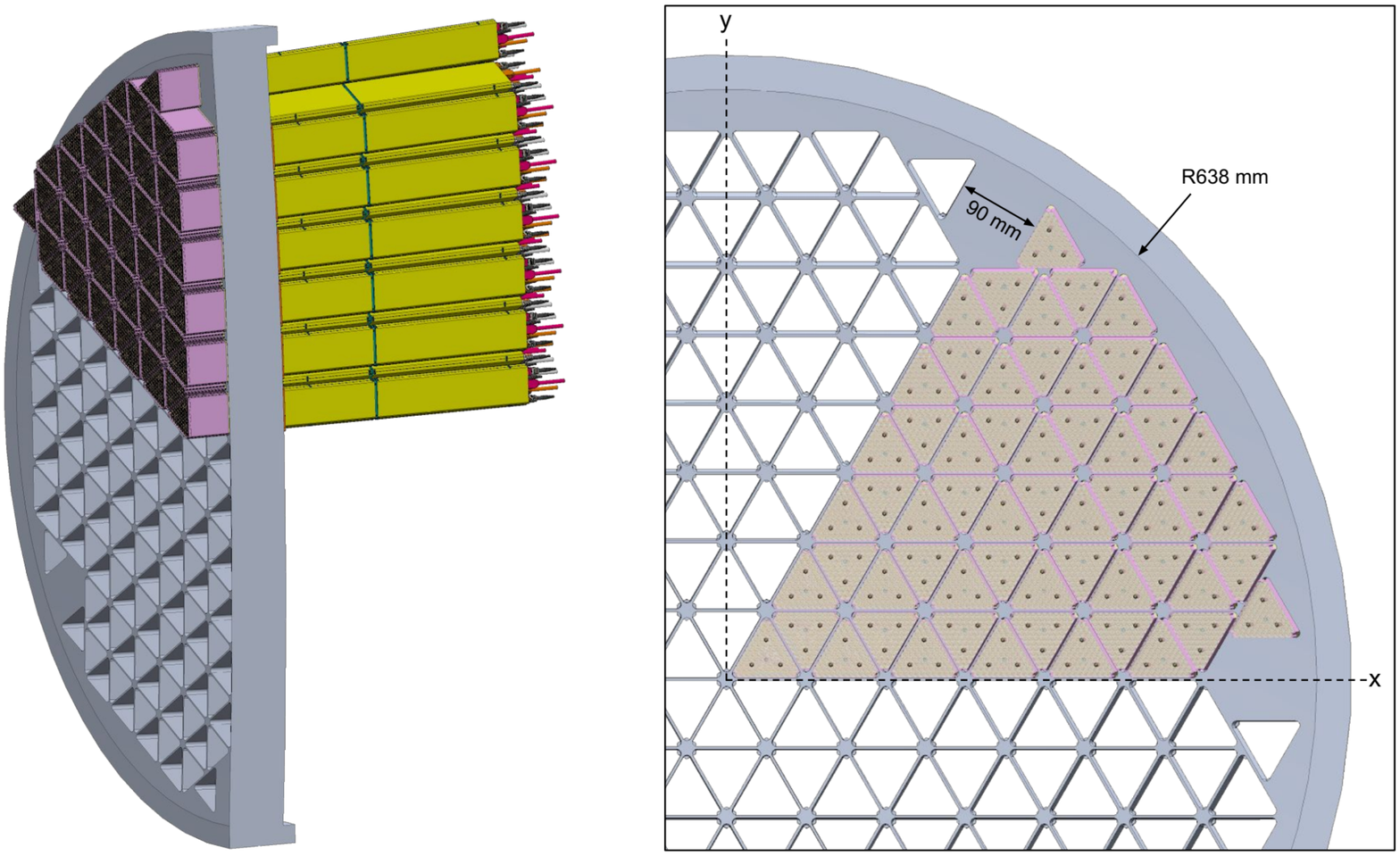}
\end{center}
\caption{Focal plane layout for 348 fiber robot rafts (with 26,100 robots) on the MegaMapper focal plane.}
\label{fig:mm-focal}
\end{figure}

\section{Instrumentation for the spectroscopic road map}

\subsection{DESI}

The Dark Energy Spectroscopic Instrument (DESI) is a 5000-fiber multi-object spectrograph that completed installation on the Kitt Peak 4-meter Mayall telescope in 2019.
The instrument met its design requirements, with
the high-throughput, 3-channel spectrographs operate at a fixed resolution of 2000 at 360 nm to 5500 at 980 nm.
After a COVID-19 shutdown for most of 2020, DESI began a 5-year, 40 million galaxy and quasar redshift survey in May 2021.
This instrument is expected to remain one of the fastest platforms for conducting wide-area redshift surveys even after the completion of this first survey in 2026, and is could be re-tasked with addressing some of the cosmology science cases listed in this white paper.

\subsection{DESI-II}

The DESI instrument is suitable for an upgrade that would double its observing power, and therefore the speed at which it could conduct the next phase of cosmology observations.
The focal plane would be upgraded from 5000 to 11,250 fiber robots using the MegaMapper 6.1-mm pitch robots currently under development.  This would consist of 150 small rafts, each with 75 robots, tiling the 812 mm-diameter focal plane (Figure \ref{fig:desi2-focal}).
These fibers would feed the existing 10 three-channel spectrographs plus an additional 6 to 9 spectrographs, with that number dependent upon the exact packing of fibers per spectrograph.
An upgrade would also consider Skipper CCDs and Skipper-enabled Front End Electronics (FEEs) for the blue cameras \cite{2020SPIE11454E..1AD}.

\subsection{Stage-5}

Several experimental designs have been proposed to advance to Stage-5 spectroscopy for cosmology experiments.
These include the MegaMapper \cite{Schlegel:2019eqc},
the Mauna Kea Spectroscopic Explorer (MSE; \cite{2019arXiv190707192M}),
and SpecTel \cite{2019arXiv190706797E}
(see Table \ref{tab_surveyspeed}).
MSE would be a new 10-m-class telescope with a 1.52 deg$^2$ field of view designed as a multi-purpose survey instrument.  For cosmology experiments, it would feature 3249 fibers feeding low/moderate resolution spectrographs from 360-950 nm.
SpecTel would be a new 10-m-class telescope with a 5 deg$^2$ field of view, also designed as a multi-purpose survey instrument. 15,000 fibers would feed spectrographs from 360-1330 nm.
The MegaMapper would be a dedicated cosmology facility intended to exploit a more cost-effective trade space, with a smaller 6.5-m telescope with a 7 deg$^2$ field of view accommodating 26,100 fibers.
The MegaMapper would re-purpose much of the DESI instrumentation.  The design uses the DESI spectrographs (and the identical SDSS-V LVM spectrographs), which have proven to be highly efficient for the purpose of redshift measurements.  The 150 fiber rafts in the DESI-II upgrade would provide 40\% of the required robots.  The investment in the DESI data systems would provide a suitable basis for the data reduction.

\begin{table}[htb]
\caption{Survey speeds for multi-fiber spectrographs as measured by the product of the telescope clear aperture, number of fibers and losses from mirror reflections.  This speed assumes a dedicated facility, which would not be possible in all cases.  Keck/FOBOS\cite{2019arXiv190707195B}, MSE\cite{2019arXiv190707192M}, SpecTel\cite{2019arXiv190706797E} and MegaMapper\cite{Schlegel:2019eqc} are proposed experiments.  LSSTspec\cite{2019arXiv190504669S,Blum22} is a notional number using MegaMapper positioners on the LSST focal plane, if optical design limitations could be overcome injecting f/1.2 light into fibers.
}
\label{tab_surveyspeed}
\begin{tabular}{lrrrrr@{.}l}\hline
Instrument (year)     & Primary/m$^2$ &  Nfiber  &  Reflections&  Product  & Speed vs& \ SDSS \\
          \hline
SDSS (1999)      &   3.68      &     640  &   0.9$^2$     &     1908  &    1&00 \\
BOSS (2009)     &   3.68      &    1000  &   0.9$^2$     &     2980  &    1&56 \\
DESI (2020)     &   9.5       &    5000  &   0.9$^1$     &   42,750  &   22&4 \\
PFS  (2023)     &  50         &    2400  &   0.9$^1$     &  108,000  &   56&6 \\
4MOST (2023)    &  12         &    1624  &   0.9$^2$     &      15,800  &    8&3 \\
{\bf DESI-Upgrade (2027)}     &   {\bf 9.5}       &    {\bf 11,250}  &   {\bf 0.9$^1$}     &   \bf{96,200}  &   {\bf50}&{\bf 4} \\
{\bf MegaMapper}&    {\bf 28}       &    {\bf 26,100}&     {\bf 0.9}$^{\bf 2}$     &  {\bf 590,000}  &  {\bf 309}&  \\
Keck/FOBOS & 77.9       &    1800  &   0.9$^3$     &  102,000  &  53&6 \\
MSE       &  78         &    3249  &   0.9$^1$     &  228,000  &  119& \\
LSSTspec  &  35.3       &    8640  &   0.9$^3$     &  222,000  &  116& \\
SpecTel   &  87.9       &   15,000 &   0.9$^2$     & 1,070,000  &  560& \\
\hline
\end{tabular}
\end{table}

\clearpage 

\section{Conclusions}

In this white paper, we present a ``Spectroscopic Road Map" from the current state-of-the art in wide-field spectroscopic mapping through the next decade.  The path outlined marks a clear trajectory from the highly successful DESI project, through its mapping-speed upgrade with DESI-II, to a Stage-5 spectroscopic program such as the MegaMapper concept.  This course enables technological development and risk retirement during the operations of DESI and DESI-II. In this approach, the Stage-5 spectroscopic program directly builds upon the instrumentation and observations of predecessor projects.

The Stage-5 program will ultimately have the capacity to map hundreds of millions of stars and galaxies.  This provides the platform for understanding the most fundamental questions in physics, such as the nature of inflation, dark energy, and dark matter, as detailed in numerous Snowmass white papers and summarized herein.  The program envisioned serves a broad range of the cosmology community and is highly complementary, if not essential, to extracting the full potential of LSST and CMB-S4, and developing line intensity mapping (LIM) as a future cosmological probe.

\appendix
\section*{Appendices}
\renewcommand{\thesubsection}{\Alph{subsection}}

\subsection{Target tables}

The appendix tabulates the redshift distributions for the DESI survey and for the DESI-II and Stage-5 samples described in Sections \ref{sec:targets-lowz}, \ref{sec:targets-midz}, \ref{sec:targets-lbg} and \ref{sec:targets-lae}.
All tables present the number of sources per square degree per redshift interval $\Delta z=0.1$.
The DESI samples include the bright galaxy sample (BGS) \cite{hahn22a}, luminous red galaxies (LRG) \cite{zhou22a}, emission line galaxies  (ELG) \cite{raichoor22a} and quasars QSO) \cite{chaussidon22a}.
The DESI-II and Stage-5 samples split the low-redshift samples into a bright sample (LOWZBRI at $r<21.6$) and a faint sample (LOWZFNT at $r<22.4$), and also include Lyman-$\alpha$ emitters (LAE) and Lyman-break galaxies (LBG).

\begin{table*}[!ht]
\small
\centering
\begin{tabular}{lccccc|cccccc}
        \hline
        \hline
        $z_{\rm min}$ & $z_{\rm max}$ & BGS & LRG & ELG & QSO & $z_{\rm min}$ & $z_{\rm max}$ & BGS & LRG & ELG & QSO\\
        \hline
        0.0 & 0.1 & 101 & 1 & 10 & 0 & 2.0 & 2.1 & 0 & 0 & 0 & 10\\
        0.1 & 0.2 & 231 & 1 & 21 & 0 & 2.1 & 2.2 & 0 & 0 & 0 & 9\\
        0.2 & 0.3 & 217 & 7 & 12 & 0 & 2.2 & 2.3 & 0 & 0 & 0 & 8\\
        0.3 & 0.4 & 97 & 31 & 6 & 1 & 2.3 & 2.4 & 0 & 0 & 0 & 7\\
        0.4 & 0.5 & 0 & 49 & 6 & 1 & 2.4 & 2.5 & 0 & 0 & 0 & 6\\
        0.5 & 0.6 & 0 & 68 & 5 & 2 & 2.5 & 2.6 & 0 & 0 & 0 & 5\\
        0.6 & 0.7 & 0 & 82 & 13 & 3 & 2.6 & 2.7 & 0 & 0 & 0 & 4\\
        0.7 & 0.8 & 0 & 95 & 54 & 5 & 2.7 & 2.8 & 0 & 0 & 0 & 4\\
        0.8 & 0.9 & 0 & 101 & 102 & 7 & 2.8 & 2.9 & 0 & 0 & 0 & 3\\
        0.9 & 1.0 & 0 & 64 & 120 & 8 & 2.9 & 3.0 & 0 & 0 & 0 & 3\\
        1.0 & 1.1 & 0 & 29 & 114 & 9 & 3.0 & 3.1 & 0 & 0 & 0 & 2\\
        1.1 & 1.2 & 0 & 10 & 108 & 10 & 3.1 & 3.2 & 0 & 0 & 0 & 2\\
        1.2 & 1.3 & 0 & 3 & 103 & 11 & 3.2 & 3.3 & 0 & 0 & 0 & 1\\
        1.3 & 1.4 & 0 & 1 & 97 & 11 & 3.3 & 3.4 & 0 & 0 & 0 & 1\\
        1.4 & 1.5 & 0 & 0 & 87 & 12 & 3.4 & 3.5 & 0 & 0 & 0 & 1\\
        1.5 & 1.6 & 0 & 0 & 55 & 12 & 3.5 & 3.6 & 0 & 0 & 0 & 1\\
        1.6 & 1.7 & 0 & 0 & 9 & 12 & 3.6 & 3.7 & 0 & 0 & 0 & 0\\
        1.7 & 1.8 & 0 & 0 & 0 & 12 & 3.7 & 3.8 & 0 & 0 & 0 & 0\\
        1.8 & 1.9 & 0 & 0 & 0 & 11 & 3.8 & 3.9 & 0 & 0 & 0 & 0\\
        1.9 & 2.0 & 0 & 0 & 0 & 11 & 3.9 & 4.0 & 0 & 0 & 0 & 0\\
        \hline
        \hline
\end{tabular}
\caption{Per-tracer redshift distribution for DESI.}
\label{tab:DESI.}
\end{table*}

\begin{table*}[!ht]
\small
\centering
\begin{tabular}{lccccc|cccccc}
        \hline
        \hline
        $z_{\rm min}$ & $z_{\rm max}$ & LOWZBRI & LOWZFNT & LAE & LBG & $z_{\rm min}$ & $z_{\rm max}$ & LOWZBRI & LOWZFNT & LAE & LBG\\
        \hline
        0.0 & 0.1 & 300 & 299 & 0 & 0 & 2.0 & 2.1 & 0 & 0 & 0 & 107\\
        0.1 & 0.2 & 545 & 237 & 0 & 0 & 2.1 & 2.2 & 0 & 0 & 0 & 116\\
        0.2 & 0.3 & 893 & 442 & 0 & 0 & 2.2 & 2.3 & 0 & 0 & 0 & 117\\
        0.3 & 0.4 & 1569 & 852 & 0 & 0 & 2.3 & 2.4 & 0 & 0 & 189 & 109\\
        0.4 & 0.5 & 967 & 673 & 0 & 0 & 2.4 & 2.5 & 0 & 0 & 153 & 94\\
        0.5 & 0.6 & 708 & 704 & 0 & 0 & 2.5 & 2.6 & 0 & 0 & 124 & 78\\
        0.6 & 0.7 & 679 & 1151 & 0 & 0 & 2.6 & 2.7 & 0 & 0 & 99 & 64\\
        0.7 & 0.8 & 372 & 885 & 0 & 0 & 2.7 & 2.8 & 0 & 0 & 80 & 56\\
        0.8 & 0.9 & 323 & 1055 & 0 & 0 & 2.8 & 2.9 & 0 & 0 & 64 & 52\\
        0.9 & 1.0 & 151 & 807 & 0 & 0 & 2.9 & 3.0 & 0 & 0 & 51 & 48\\
        1.0 & 1.1 & 34 & 256 & 0 & 0 & 3.0 & 3.1 & 0 & 0 & 40 & 43\\
        1.1 & 1.2 & 24 & 226 & 0 & 1 & 3.1 & 3.2 & 0 & 0 & 0 & 35\\
        1.2 & 1.3 & 17 & 123 & 0 & 2 & 3.2 & 3.3 & 0 & 0 & 0 & 25\\
        1.3 & 1.4 & 2 & 28 & 0 & 5 & 3.3 & 3.4 & 0 & 0 & 0 & 15\\
        1.4 & 1.5 & 0 & 0 & 0 & 10 & 3.4 & 3.5 & 0 & 0 & 0 & 9\\
        1.5 & 1.6 & 0 & 0 & 0 & 19 & 3.5 & 3.6 & 0 & 0 & 0 & 4\\
        1.6 & 1.7 & 0 & 0 & 0 & 32 & 3.6 & 3.7 & 0 & 0 & 0 & 2\\
        1.7 & 1.8 & 0 & 0 & 0 & 49 & 3.7 & 3.8 & 0 & 0 & 0 & 1\\
        1.8 & 1.9 & 0 & 0 & 0 & 69 & 3.8 & 3.9 & 0 & 0 & 0 & 0\\
        1.9 & 2.0 & 0 & 0 & 0 & 90 & 3.9 & 4.0 & 0 & 0 & 0 & 0\\
        \hline
        \hline
\end{tabular}
\caption{Per-tracer redshift distribution for DESI-II.}
\label{tab:DESI2.}
\end{table*}

\begin{table*}[!ht]
\small
\centering
\begin{tabular}{lcccccc|ccccccc}
        \hline
        \hline
        $z_{\rm min}$ & $z_{\rm max}$ & LOWZBRI & LOWZFNT & ELG & LAE & LBG & $z_{\rm min}$ & $z_{\rm max}$ & LOWZBRI & LOWZFNT & ELG & LAE & LBG\\
        \hline
        0.0 & 0.1 & 300 & 299 & 0 & 0 & 0 & 3.0 & 3.1 & 0 & 0 & 0 & 197 & 191\\
        0.1 & 0.2 & 545 & 237 & 0 & 0 & 0 & 3.1 & 3.2 & 0 & 0 & 0 & 168 & 161\\
        0.2 & 0.3 & 893 & 442 & 0 & 0 & 0 & 3.2 & 3.3 & 0 & 0 & 0 & 142 & 123\\
        0.3 & 0.4 & 1569 & 852 & 0 & 0 & 0 & 3.3 & 3.4 & 0 & 0 & 0 & 120 & 88\\
        0.4 & 0.5 & 967 & 673 & 0 & 0 & 0 & 3.4 & 3.5 & 0 & 0 & 0 & 102 & 66\\
        0.5 & 0.6 & 708 & 704 & 0 & 0 & 0 & 3.5 & 3.6 & 0 & 0 & 0 & 0 & 57\\
        0.6 & 0.7 & 679 & 1151 & 0 & 0 & 0 & 3.6 & 3.7 & 0 & 0 & 0 & 0 & 56\\
        0.7 & 0.8 & 372 & 885 & 0 & 0 & 0 & 3.7 & 3.8 & 0 & 0 & 0 & 0 & 57\\
        0.8 & 0.9 & 323 & 1055 & 0 & 0 & 0 & 3.8 & 3.9 & 0 & 0 & 0 & 0 & 55\\
        0.9 & 1.0 & 151 & 807 & 0 & 0 & 0 & 3.9 & 4.0 & 0 & 0 & 0 & 0 & 48\\
        1.0 & 1.1 & 34 & 256 & 0 & 0 & 1 & 4.0 & 4.1 & 0 & 0 & 0 & 0 & 38\\
        1.1 & 1.2 & 24 & 226 & 0 & 0 & 3 & 4.1 & 4.2 & 0 & 0 & 0 & 0 & 28\\
        1.2 & 1.3 & 17 & 123 & 0 & 0 & 7 & 4.2 & 4.3 & 0 & 0 & 0 & 0 & 18\\
        1.3 & 1.4 & 2 & 28 & 0 & 0 & 15 & 4.3 & 4.4 & 0 & 0 & 0 & 0 & 12\\
        1.4 & 1.5 & 0 & 0 & 171 & 0 & 29 & 4.4 & 4.5 & 0 & 0 & 0 & 0 & 8\\
        1.5 & 1.6 & 0 & 0 & 160 & 0 & 54 & 4.5 & 4.6 & 0 & 0 & 0 & 0 & 8\\
        1.6 & 1.7 & 0 & 0 & 146 & 0 & 90 & 4.6 & 4.7 & 0 & 0 & 0 & 0 & 8\\
        1.7 & 1.8 & 0 & 0 & 174 & 0 & 139 & 4.7 & 4.8 & 0 & 0 & 0 & 0 & 9\\
        1.8 & 1.9 & 0 & 0 & 138 & 0 & 197 & 4.8 & 4.9 & 0 & 0 & 0 & 0 & 9\\
        1.9 & 2.0 & 0 & 0 & 116 & 0 & 255 & 4.9 & 5.0 & 0 & 0 & 0 & 0 & 9\\
        2.0 & 2.1 & 0 & 0 & 73 & 0 & 304 & 5.0 & 5.1 & 0 & 0 & 0 & 0 & 9\\
        2.1 & 2.2 & 0 & 0 & 21 & 818 & 332 & 5.1 & 5.2 & 0 & 0 & 0 & 0 & 7\\
        2.2 & 2.3 & 0 & 0 & 0 & 701 & 334 & 5.2 & 5.3 & 0 & 0 & 0 & 0 & 6\\
        2.3 & 2.4 & 0 & 0 & 0 & 600 & 312 & 5.3 & 5.4 & 0 & 0 & 0 & 0 & 4\\
        2.4 & 2.5 & 0 & 0 & 0 & 513 & 274 & 5.4 & 5.5 & 0 & 0 & 0 & 0 & 3\\
        2.5 & 2.6 & 0 & 0 & 0 & 439 & 236 & 5.5 & 5.6 & 0 & 0 & 0 & 0 & 2\\
        2.6 & 2.7 & 0 & 0 & 0 & 375 & 210 & 5.6 & 5.7 & 0 & 0 & 0 & 0 & 1\\
        2.7 & 2.8 & 0 & 0 & 0 & 320 & 201 & 5.7 & 5.8 & 0 & 0 & 0 & 0 & 0\\
        2.8 & 2.9 & 0 & 0 & 0 & 272 & 204 & 5.8 & 5.9 & 0 & 0 & 0 & 0 & 0\\
        2.9 & 3.0 & 0 & 0 & 0 & 232 & 204 & 5.9 & 6.0 & 0 & 0 & 0 & 0 & 0\\
        \hline
        \hline
\end{tabular}
\caption{Per-tracer redshift distribution for MegaMapper.}
\label{tab:MegaMapper.}
\end{table*}

\clearpage
\newpage
\mbox{~}
\clearpage
\newpage







\newpage
\bibliographystyle{utphys}
\bibliography{main}

\end{document}